\documentclass[aps,amssymb,twocolumn]{revtex4}
\usepackage{graphics}

\def\d {{\rm d}}

\def\dpl{\gamma}
\def\ddpl{\dot\dpl}

\begin{document}
\title {A dynamical approach to glassy materials}
\author{Ana\"el Lema\^{\i}tre}
\affiliation{Department of physics, University of California, Santa Barbara, 
California 93106, U.S.A.}
\affiliation{CEA --- Service de Physique de l'\'Etat Condens\'e,
Centre d'\'Etudes de Saclay, 91191 Gif-sur-Yvette, France}
\date{\today}
\begin{abstract}
Typical properties of glassy materials are shown to be captured by a mean-field
free-volume theory. Relaxation processes are supposed to be free-volume activated,
and different entropy barriers are associated with density relaxation
and shear motion.
Free-volume time logarithmic relaxation, Kohlrausch-Williams-Watts,
and power law viscosity result from
the non-linear dynamics of spatially averaged quantities.
The exponents associated with these phenomena are related to a single parameter
of the theory.
The theory also accounts for coexistence of jamming transitions and non-linear rheology.
\end{abstract}

\maketitle

\section{Introduction}


The emergence of slow modes of relaxations is a major signature 
of the glassy behavior.~\cite{goetze92}
The first consequence of large equilibration time is that the properties 
of a glass evolve on experimental timescales, the system ages.
Despite the huge amount of work that has been devoted to glassy systems,
a general understanding of slow relaxation processes and of their
relation to macroscopic rheology is still lacking.
For example, experimental and numerical data converge to a stretched exponential 
shape of relaxation phenomena, the so-called Kohlrausch-Williams-Watts
(KWW) relaxation,~\cite{ediger96,larson99,angell00}
while mode coupling theory predicts power law behavior.~\cite{goetze92}
When it comes to extend these results to non-linear rheology, 
the situation is even more difficult from the theoretical point of view,
since the system is driven far from equilibrium.
At an applied strain rate, power law viscosities or 
the emergence of a yield stress, are common features 
of complex fluids.~\cite{larson99} 
MD simulations of Lennard-Jones systems~\cite{berthier01,yamamoto97,yamamoto98}
and recent experiments on colloidal glasses~\cite{bonn02a,bonn02b}
are also consistent with such properties.

Recent theoretical approaches to non-linear
rheology~\cite{sollich97,sollich98,hebraux98,berthier00,fielding00} 
have called on ideas originating in mode coupling theory, 
or trap models of spin glasses.~\cite{goetze92,bouchaud96,monthus96}
These approaches, however, fail to provide a description of non-linear
rheology that is consistent with experimental observations;~\cite{bonn02a,bonn02b}
moreover, these models, based only on a phase-space picture of glassiness, 
do not lead to a clear identification of real-space physical mechanism 
that are responsible for an observed macroscopic behavior.
Some other attempts have been directed towards phenomenological 
approaches,~\cite{derec01} but those are still unsatisfactory, 
and do not offer a microscopic picture, even heuristic, of the undergoing physical 
mechanisms.

The current work lies at the convergence point of several recent theoretical
ideas in various fields.
Firstly, it is based on the idea by Liu, Nagel and coworkers, 
that some unique mechanism lies 
behind the profound similarities displayed by structural glassy materials,
be they colloidal glasses, granular materials, foams,\dots~\cite{liu98}
Drawing on this idea, the fundemental assumption underlying the
current work is that macroscopic 
properties of dense materials results from major {\it dynamical} properties 
of structural rearrangement of the contact network.

Another important element of the current work comes from the study 
of density relaxation 
in granular materials.~\cite{knight95,boutreux97,nowak98}
Under vertical tapping, granular materials present a time-logarithmic relaxation 
of the occupied volume; this is in strong analogy with free-volume relaxation, 
commonly observed in glasses, and of central importance for aging.~\cite{larson99}
However, in the case of tapped granular materials, 
density relaxation cannot result from thermal activation
since there is no thermal bath:
the logarithmic density relaxation has been
explained by introducing an equation of motion for free-volume,
which characterizes the internal state of the granular material.

Finally, this work uses a theory introduced recently by Falk and Langer
to account for elasto-plastic transition in amorphous solids.~\cite{falk98,falk00} 
The so-called shear transformation zone (STZ) theory, 
originates from ideas by Argon, Spaepen and others
to describe creep in metallic
alloys;~\cite{spaepen77,argon79a,argon79b,spaepen81,argon83}
it has been primarily developed in the framework of fracture mechanics,
and provides a general scheme for jamming as resulting from a structuration 
of the material by creation of arrangements oriented along principal directions
of an applied stress. 
STZ theory calls to free-volume arguments, but in fact,
free-volume remains a parameter and enters constants of the theory.

These ideas lead me to propose a general approach to the rheology of
dense materials, leading to a simple set of constitutive equations.~\cite{lemaitre01b}
The purpose of the this article is to  elaborate the theoretical arguments 
underlying these equations a to detail some of their consequences.

Two types of internal state variables are used: free-volume, 
which is an isotropic property of a molecular packing, and populations of arrangements, 
which are related to the distortion of the contact network between molecules  
(hence introduce anisotropy).
Those variables evolves dynamically by free-volume activated processes:
two types of rearrangements are associated to compaction and shear motion.
Free-volume enters activation factors that control both types 
of elementary rearrangements, and it is also a dynamical quantity. 

KWW relaxation has been shown to result from 
the existence of a distribution of timescales;~\cite{larson99,angell00}
and it seems that this is often thought to be a necessary requirement.
The first contribution of the current work is to show that it is not necessary,
and that KWW relaxation also result from a more simple mechanism:
the interplay between free-volume logarithmic relaxation and
free-volume activated shear deformation.

The second contribution of this work is to establish a relation 
between KWW relaxation, time-logarithmic density relaxation, 
and power law viscosity of a strongly sheared glassy material.
A single parameter of the theory determines these three types of phenomena.
Power law viscosity will be shown to result from 
the existence of shear induced dilatancy. 
I propose one of the simplest set of equations that can account 
for those three properties is a single framework.

The third contribution of this work is to present a theoretical framework
which accounts for two coexisting mechanisms for jamming.
In mode-coupling theory, jamming results from an entropy crisis:
at low temperature, the system supports any applied stress because 
no motion of molecules is allowed.
Resistance to shear is therefore an intrinsic property of a material,
determined by the thermodynamic parameters of the system,
and results from a phase transition.
In STZ theory, jamming results from a structuration 
of the contact network:
jamming requires some amount of plastic deformation
to induce the creation of anisotropic structures that 
can support a static stress.
Those two mechanisms are shown here to be incorporated in a unique theory. 
The difference between hard and soft glassy materials
will be identified with the cases when one or the other process dominates.
The interplay between those two mechanisms
lead to a novel interpretation of brittleness and ductility,
and provide age-dependent dynamical yield criteria.

\section{Constitutive equations}
\subsection{Preliminary settings}

The main assumption made in this work is that the out-of-equilibrium
dynamics of a glass can be modeled in the framework of a mean-field approximation.
By mean-field, I mean that spatially averaged quantities account 
for the local state of the material. 
Constitutive equations are proposed that relate the average shear stress,
to the average shear strain.
The internal state of the material is determined by free-volume
and by densities of arrangements (to be defined further).


The forces in the material are supposed to be characterized by a stress tensor, 
which is written
$$
\left(
\matrix{ -P& \sigma\cr
\sigma &-P\cr}
\right)
$$
where $P$ is the pressure, and $\sigma$ the shear stress.
The deformation tensor is defined accordingly, and composed of shear,
and of isotropic deformations, directly related to dilatancy.
The average free-volume per molecule is denoted $v_f$,
and is related to the total volume $V$ by, 
$$
V=N v_f + N v^{\rm rcp}
\quad,
$$
where $v^{\rm rcp}$ is the volume per molecule of the material in random close packing,
and $N$ is the number of molecules.

Under strong shear, soft glassy materials, like clays, foams, or granular materials, 
can achieve large deformations; the evolution of the system
is more similar to the flow of a liquid than to the deformation of a solid.
In this work I present a theory of deformation that is expected to hold
for large deformations of amorphous solids or soft glassy materials.
Since there is a long-standing matter of debate in the mechanical community
about linear versus non-linear deformations,~\cite{lubliner90}
it seems necessary to clarify this issue.
I show here that the notion of inherent states~\cite{stillinger84} 
offers a simple physical interpretation of small and large deformations.

At low temperature, a supercooled liquid or a glassy material evolves in a complex
energy landscape which can be partitioned in domains of influence of local minima.
For almost all geometrical configuration of the molecules, a unique inherent state 
is defined as the local minimum to which the systems relaxes
if suddenly quenched to zero temperature,~\cite{stillinger84} 
and in the absence of forcing.
When a material is forced, a constant stress or strain pulls the system
through the energy landscape.
Each geometrical configuration reached lies in the domain associated with
an instantaneous inherent state.
The time series of instantaneous inherent states provides a coarse-grained
description of the deformation of the material, and is taken here as the definition
of irreversible, plastic, deformation.

The instantaneous inherent state is a natural reference state for 
a given molecular configuration, and deviation from the instantaneous inherent state 
defines an {\it inherent (shear) deformation} $\epsilon^{\rm in}$.
Starting from a known initial configuration, the total shear deformation, 
$\epsilon= \epsilon^{\rm in} + \gamma$, results both from inherent deformation, 
and from the {\it flow} of inherent states in the phase space.
By definition, the deformation $\gamma$ is measured between two inherent states:
it is not a state variable, since it is defined only from the knowledge 
of a configuration designed as ``initial''. $\epsilon^{\rm in}$, however, is a state
variable, since it is well defined for any instantaneous configuration of the system.
Moreover, $\epsilon^{\rm in}$ is expected to be small, since it
measures the deformation between two nearby configurations in phase space;
elasticity results from a harmonic approximation for every local minimum:
it permits to write the stress as given by a Hooke law, 
$\sigma=\mu\,\epsilon^{\rm in}$. 
The equation that governs $\sigma$ can finally be written:
\begin{equation}
\dot\sigma = \mu\,(\dot\epsilon-\dot\gamma)
\quad.
\label{eqn:sigma:0}
\end{equation}
The rate of plastic deformation $\dot\gamma$ will come out of a statistical analysis
of hopping motion in the phase-space. 
It will be written as a function of the state variables of the system, 
$\sigma$, $v_f$, an others, to define the constitutive law of the material.

\subsection{Assumptions}
\subsubsection{Free-volume}

This work relies on the idea that, in dense materials,
macroscopic deformation results from free-volume activated rearrangements.
This is expected because temperature is so low that thermal activation 
becomes irrelevant, and activated processes are controlled by entropic fluctuations
of the free-volume.
A rearrangements occurs if sufficient space exists in the neighborhood
of a given point; the transition probabilities are thus
determined by the size distribution of voids.
In order to work at a mean-field level, the volume distribution is characterized
by the single scalar value $v_f$; in order to write activation factors,
it is necessary to provide an Ansatz for the distribution of voids in the material.

In the original works by Cohen and Turnbull on free-volume theory,
it was argued that the voids in a material are given by 
a Poisson distribution.~\cite{cohen59,turnbull61,turnbull70}
In their picture, the motion of a molecule in a glassy material is permitted only 
if it can move into a hole of size, say $v_1$.
The probability to find a hole larger than $v_1$, is proportional to $\exp(-v_1/v_f)$:
this leads to estimate a diffusion constant $D\propto\exp(-v_1/v_f)$,
which is a basis of free-volume approaches. 
Important differences, however, exist between various 
uses and interpretations of free-volume theory.

Falk and Langer use free-volume activation factors in the definition 
of STZ theory.~\cite{falk00}
Their argument borrowed from original ideas by Edwards and coworkers
for granular materials:~\cite{edwards89,mehta89,edwards94} 
the only extensive variable is supposed to be the volume $V$;
the number of states available to the system
is roughly proportional to $(v_f/h)^N$, where $h$ is an arbitrary constant
with dimension of a volume; the entropy is defined as,
$$
S(V,N) \simeq N\,\ln\left({v_f\over h}\right) = N\,\ln\left({V- N v^{\rm rcp}\over N h}\right)
$$
and the intensive quantity, analogous to temperature, is $\chi$:
$$
{1\over\chi}\equiv {\partial S\over\partial V} = {1\over v_f}
\quad.
$$
Free-volume thus enters activation factors of the form $\exp(-v_1/v_f)$.

In STZ theory, this discussion turns out to be somewhat formal
since free-volume is taken constant:
$v_f$ is absorbed in the phenomenological constants of the resulting equations.
It also leads to complicated expressions for transition rates as functions
of the applied stress (because $v_1$ is supposed to be a function of $\sigma$).

However, free-volume should vary when the material dilates or contracts.
This idea has been considered recently by several authors in the analysis 
of density relaxation of granular materials subjected 
to vertical tapping.~\cite{knight95,boutreux97,nowak98}
In these works, free-volume is understood as a purely dynamical quantity,
and relaxes by the motion of single grains in holes created by volume fluctuations.
A Poisson distribution of volume fluctuations is assumed at all times:
this leads to an equation of motion of the form,
\begin{equation}
\dot v_f = -R_1\,\exp\left[{-{v_1\over v_f}}\right]
\quad.
\label{eqn:vf:0}
\end{equation}
The activation factor $\exp\left[{-{v_1/v_f}}\right]$ is the probability that
a hole of size larger than $v_1$ is present at a given point in the material;
$R_1$ is determined by the tapping frequency.

The current work borrows from those different lines of thought.
Drawing on the works initiated on granular compaction, free-volume is a dynamical 
quantity; voids are expected to redistribute fast enough in the material;
fast means, faster than free-volume activated collective rearrangements themselves.
This is justified because redistribution of free-volume requires very tiny 
displacements of the molecules, and voids, considered as 
particles, are expected to diffuse faster than molecules themselves:
around a void, there is by definition an excess of free-volume.

There is an important difference, however, from the picture used 
in~\cite{knight95,boutreux97,nowak98}
where compaction results from the motion of a single grain in a hole.
Here, borrowing on the argument of STZ theory,
compaction results from elementary rearrangements, involving several
molecules at a mesoscopic scale. 
Probabilities of transitions are estimated by entropic arguments.
Due to the redistribution of voids, the number of states available 
to the system is supposed to be proportional to $(v_f/h)^N$ at all times:
the fast redistribution of voids allows the system to 
realize a quasi-equilibrium thermodynamic ensemble, 
determined by an intensive variable, $v_f$.

The existence of an intensive variable, associated to 
the fluctuations of molecular configurations, is supported
by the observation of an effective fluctuation-dissipation theorem
in sheared fluids.~\cite{berthier01}
For reasons of clarity, and to be specific, 
I reserve the word temperature for the intensive quantity associated
with the fluctuations of energy, or for the specific kinetic energy;
therefore, I will not comply with to the recent use of ``effective temperature''
for something that is neither thermodynamic nor granular temperature.
Nevertheless, the observation of an effective fluctuation-dissipation theorem
indicates that such an intensive quantity exists, without identifying 
what this quantity actually is.
Here, it is identified to $v_f$ and is associated with entropic fluctuations of 
molecular configurations at a mesoscopic scale.

The important novelty introduced in the current work it that 
$v_f$ enters activation factors and is also a dynamical quantity.
This is in weak analogy with granular temperature, which is a temperature, 
and evolves dynamically.
But granular temperature couples with properties of the material only as far as
it determines a collision frequency,~\cite{savage79,lemaitre01a}
while free-volume enters exponential prefactors in transformation rates.
Another important element of the current theory is that the dynamics of 
free-volume results both from density relaxation by activated rearrangements 
-- which exists in the absence of forcing --
and from dilatancy induced by the macroscopic shear flow 
-- which occurs when the system is driven out-of-equilibrium.
The non-linear coupling with the mean-flow will be addressed later;
I will first continue the description of rearrangement processes.

\subsubsection{Two types of activated processes}
The typical elementary compaction process,
corresponds to the system going through a saddle point in phase space.
The rate of such transformations is controlled by free-volume fluctuations 
and this saddle is characterized by a typical activation volume $v_1$,
which depends on the shape of the molecules, 
and on the details of microscopic interactions.
The activated hopping of the system through this type of saddle  
permits density relaxation, 
and leads to an equation of motion for $v_f$ of the form~(\ref{eqn:vf:0}).

Other types of elementary rearrangements occur in the material, 
in particular leading to elementary shear motion.
Before presenting a more elaborate scheme for macroscopic shear deformation,
let me use a very minimal argument:
shear strain is supposed to be proportional to shear stress: $\dot\gamma = D\sigma$;
where $D$ is proportional to a free-volume activation factor.

The current approach relies on the remark that the relative motion of molecules 
that permit compaction and shear are different;
different types of collective rearrangements,
should require different activation volumes.
In phase space, shear and compaction correspond to two types of saddle points,
and those saddle points are entropic barriers of different heights.
The activation volume for an elementary shear is denoted $v_0$, 
and {\it a priori} differs from $v_1$. This leads to
\begin{equation}
\dot\gamma \propto \exp\left[{-{v_0\over v_f}}\right]\,\sigma
\quad.
\label{eqn:dotgamma:0}
\end{equation}

One major purpose of this work is to study the consequences of
introducing complexity, not through a non-trivial distribution of 
timescales,~\cite{angell00,schlesinger84}
but through the existence of two types of entropy barriers of unequal heights.
The set of equations~(\ref{eqn:sigma:0}),~(\ref{eqn:vf:0}), and~(\ref{eqn:dotgamma:0}) 
will be shown to account for logarithmic relaxation of $v_f$ 
and for a KWW relaxation of the stress after a strain increment.

\subsubsection{Shear deformation}

\paragraph{Shear transformation zones.}
Equation~(\ref{eqn:dotgamma:0}) accounts for a free-volume dependent viscosity:
the material described is a liquid (in fact a non-Newtonian liquid).
It is however, a very rough description of shear motion.

The theoretical description of shear deformation that I present
is directly inspired by STZ theory.~\cite{falk98,falk00}
The first assumption on which STZ theory is based is that 
elementary shear deformation does not occur just anywhere in the material, 
but in zones, where a rearrangement is permitted by the local configuration of 
molecules.
A shear transformation zone (STZ) is thus defined as a locus within a material,
where an elementary shear is possible.
Another essential assumption is that STZ's are two-state systems. 
This can be understood as follows:
An elementary arrangement corresponds to the opening and closing of some 
contacts between molecules; once such a rearrangement has occurred 
somewhere in the material, these molecules cannot shear further in the same direction, but they can shear backwards.
A symmetry is thus introduced by shearing motion, and (at least) two types 
of arrangements can be identified that are transformed into one another,
as pictured here:
\begin{center}
\unitlength = 0.0011\textwidth
\begin{picture}(100,60)(0,0)
\put(50,40){\makebox(0,0){\large$R_+$}}
\put(50,0){\makebox(0,0){\large$R_-$}}
\put(-50,0){\resizebox{200\unitlength}{!}{\includegraphics{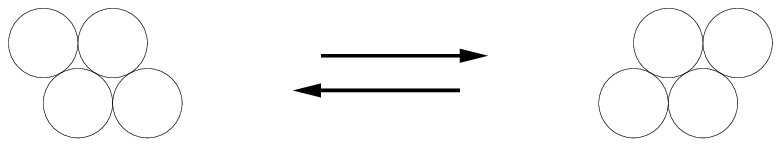}}}
\end{picture}
\end{center}
Note that an STZ is not supposed to contain only four molecules, 
but it is a mesoscopic object that accounts for the structure of several 
neighboring molecules (say of order 10).
To simplify the analysis, only one pair of types of arrangements is considered,
aligned along the principal directions of the stress tensor.
Macroscopic motion results from the statistics of elementary rearrangements,
and reads:
\begin{equation}
\dot\gamma = {\cal A}_0\,(R_+\,n_+-R_-\,n_-)
\label{eqn:stz:0}
\end{equation}
where $n_\pm$ denote the number density of arrangements, and where
$R_\pm$ denote the rate of transition $\pm\to\mp$.

Those rates result from activation processes determined by free-volume, 
but also by force fluctuations. 
A STZ shears if there is sufficient free-volume at its location,
and if the local bias of the force network triggers the shear in the appropriate 
direction.
In this work, force and volume fluctuations are supposed to be uncorrelated;
the probabilities associated with those two types of fluctuations factorize:
$R_\pm = R^v(v_f) R_\pm^\sigma(\sigma)$.
From the preceding discussion, $R^v(v_f) = \exp(-v_0/v_f)$.
In order to estimate the factors $R_\pm^\sigma$ associated with the fluctuations
of the force network, let me first remark that $R_\pm^\sigma(\sigma)$ is a positive, 
increasing function of $\sigma$.
Moreover, an elementary shear is expected to be triggered by {\em large forces}.
The distributions of static forces in glassy systems have been shown to 
share strong similarities with the force distribution in
a granular material,~\cite{ohern01} in which case, large forces are 
distributed exponentially.~\cite{coppersmith96,radjai98a,radjai98b}
This justifies an exponential dependency of $R_\pm^\sigma$ on $\sigma$:
$$
R_\pm = R_0\,\exp\left[{-{v_0\over v_f}}\right]\,\exp\left[{\pm{\sigma\over\bar\mu}}\right]
\quad,
$$
with $R_0$, the update frequency of microscopic processes.
Note however, that the exponential form for the activation factors $R_\pm^\sigma$
plays no role in the following, since these factors will be linearized.
The parameter $\bar\mu$ measures the typical stress that must be overcome 
to trigger a rearrangement. 
In hard-sphere systems, it is expected to be proportional
to the pressure $P$. In general, $\bar\mu$ is essentially the energy
required to break a bond between two molecules, times the density of bonds. 
Here, $\bar\mu$ will be taken constant, either because $P$ fixed, 
or because it depends on some interaction potential.
At some point, $\bar\mu$ will be set to unity, which fixes the unit of forces.

\paragraph{Dynamics of arrangements.}

Following~\cite{falk98,falk00}, the equation of motion for the populations $n_\pm$ is written,
\begin{equation}
\label{eqn:stz:npm}
\dot n_\pm = R_\mp n_\mp - R_\pm n_\pm +\sigma\,\ddpl\,({\cal A}_{c} - {\cal A}_{a}\,n_\pm)
\quad.
\end{equation}
The first two terms of the rhs account for the transitions between both
types of STZ. The last term accounts for the renewal of 
molecular configuration by the macroscopic flow:
STZ constantly appear and disappear during plastic deformation.
The renewal of configurations is supposed to be proportional to 
the work of plastic deformation, $\sigma\,\ddpl$, which 
is thus a common factor of creation and annhilitation terms.

\paragraph{Shear-induced dilatancy.}
In this work, the dynamics of rearrangements is coupled to free-volume dynamics.
Part of this coupling results from the factor $R^v(v_f)$ that controls
shear transformation, but the macroscopic flow is also expected
to influence the relaxation of free-volume.

Free-volume dynamics is derived in analogy with the dynamics 
of the populations $n_\pm$:
plastic deformations corresponds to the evolution of the system in the phase-space
along trajectories that do not necessarily permit a minimization of free-volume.
In real space, plastic deformation requires the constant renewal of local
arrangements, and new arrangements are local configurations of large entropy;
by definition, their density is not optimized. 
Therefore, the renewal of configurations by the shear flow leads to dilatancy.
In experiments on suspensions, for example, the system is fluidized by 
applying a strong shear which rejuvenates the glassy structure.~\cite{cloitre00}

In order to quantify the average dilatancy induced by the shear flow,
I assume that a fraction of the work of plastic deformation, $\sigma\,\dot\gamma$
in used for enthalpy production, $P \dot v_f$.
The equation for $v_f$ is:
\begin{equation}
\dot v_f = - R_1\,\exp\left[-{v_1\over v_f}\right] + {{\cal A}_v\over P}\,\sigma\,\dot\gamma
\label{eqn:vf:1}
\end{equation}
This equation is an essential element of the current theory. 
The non-linear coupling with the macroscopic flow shows up when the system 
is strongly driven out-of-equilibrium.
The competition between self-relaxation
and shear induced dilatancy
will be shown to lead to a power law viscosity.

\subsection{Equations of motion}

From~(\ref{eqn:stz:npm}), the equation of motion for the total density, 
$n_T=n_++n_-$, of arrangements reads,
$$
\dot n_T = \sigma\,\ddpl\,( 2\,{\cal A}_{c} - {\cal A}_{a}\,n_T)
\quad.
$$
The creation and destruction terms account for an entropic mixing 
of molecular configurations. 
The previous equation shows that this mixing leads to a convergence of
$n_T$ to an asymptotic value, $n_\infty=2\,{\cal A}_{c}/{\cal A}_{a}$;
therefore, $n_T=n_\infty$ defines a state of maximal entropy 
of molecular configurations.
For this reason, I do not expect $n_T$ to be different from $n_\infty$ unless
in very special circumstances, {\it e.g.} if some level of crystallization
exists in the material.
In particular, an amorphous solid is usually produced by a rapid quench
of a liquid at low temperature: the system comes from a state of high entropy,
therefore, in the resulting glass, $n_T$ is expected to achieve its asymptotic value.
In this work, I assume that $n_T=n_\infty$ at all times.
The dynamics of local arrangements is thus determined only by the bias $n_--n_+$
between the populations of STZ.

Equations~(\ref{eqn:stz:0}) and~(\ref{eqn:stz:npm})
are written in a more suitable form by introducing the variable
$$
\Delta = {n_--n_+\over n_\infty}
\quad,
$$
and parameters,
$\epsilon_0 = {{\cal A}_0\,{\cal A}_c/{\cal A}_a}$,
$\mu_0 = {\cal A}_0\,{\cal A}_c$, and $E_0 = 2\epsilon_0\,R_0$.
It is also convenient to introduce $\kappa=v_1/v_0$, $E_1=R_1/v_0$
and $\alpha = {\cal A}_v/(P v_0)$, and to rescale free-volume as,
$$
\chi = {v_f\over v_0}
\quad.
$$

The complete set of constitutive equations is finally:
\begin{eqnarray}
\label{eqn:stzdil:1}
\ddpl&=&E_0\,\exp\left[{-{1\over\chi}}\right]\,
\left(\sinh\left({\sigma\over\bar\mu}\right)-\Delta\,\cosh\left({\sigma\over\bar\mu}\right)\right)\\
\label{eqn:stzdil:2}
\dot\Delta&=&{\ddpl\over\epsilon_0}\,\left(1- \mu_0\,\sigma\,\Delta\right)\\
\label{eqn:stzdil:3}
\dot\chi &=& -E_1 \exp\left[-{\kappa\over\chi}\right] + \alpha\,\sigma\,\ddpl
\quad.
\end{eqnarray}
Those equations involve two state variables, $\Delta$ and $\chi$ which
play very different roles.
$\Delta$ evolves only when the material is sheared, therefore accounts
for some sort of structural memory induced by deformation imposed on the material.
On the contrary, $\chi$ evolves spontaneously, and relaxes towards 0.
Since this relaxation is slow, $\chi$  also encodes some sort of memory, 
but a very different type of memory than $\Delta$, 
a memory which constantly evolves as the system ages.

Note that the asymptotic value for the relaxation of $\chi$ (which here is 0)
may, in fact, be temperature dependent;
this question is eluded in the current work:
it seems a very reasonable assumption for materials like suspensions 
if not for glasses below the glass transition, because the logarithmic
slow-down of density relaxation around $\kappa$ prevents it to reach 
asymptotic values far below ($\chi_\infty<<\kappa$).
The parameters involved in the theory should depend on pressure and temperature,
and this dependency results from particularities of a given material;
for example the update frequencies $R_1$ and $R_0$ are expected to 
be proportional to $\sqrt{T}$ (where $T$ is the temperature)
in a dense hard-sphere material.~\cite{lemaitre01a}
Such questions, however, are not the focus of this study;
in this article, the values of the parameters are constant and can 
be understood as a given thermodynamical state of the system. 

\section{Interlude}
Equations~(\ref{eqn:stzdil:1}-\ref{eqn:stzdil:3}) present non-linear couplings
between shear motion and density relaxation.
Before studying the complete set of equations in various tests,
I start with two simple cases, when one or the other process dominates.
I show here that 
jamming transitions are captured by equations~(\ref{eqn:stzdil:1}-\ref{eqn:stzdil:2}),
and free-volume relaxation by equation~(\ref{eqn:stzdil:3}).

\subsection{Free-volume relaxation} 
In the absence of shear stress, $\sigma=0$ and $\dot\gamma=0$;
the system~(\ref{eqn:stzdil:1}-\ref{eqn:stzdil:3}) reduces to the single equation:
$$
\dot \chi = -E_1 \exp\left[-{\kappa\over\chi}\right]
\quad,
$$
which accounts for density relaxation.

At short time, this process is dominated by the initial value, $\chi_0$:
$$
\dot \chi \simeq -E_1 \exp\left[-{\kappa\over\chi_0}\right]
\quad;
$$
this defines a time scale for short-time free-volume relaxation:
$$
\tau_\chi = {\chi_0\over E_1}\;\exp\left[{\kappa\over\chi_0}\right]
\quad.
$$

The long time behavior is estimated from the integral expression:
$$
E_1 t = \int_{\chi(t)}^{\chi(0)} \;\exp\left[{\kappa \over \chi}\right] \d \chi \simeq 
\kappa\,\exp\left[{\kappa\over \chi(t)}\right]
$$
At long time, the integral is dominated by the small values of $\chi$, 
{\it i.e.} by the ultimate value reached, $\chi(t)$. Therefore,
\begin{equation}
\chi(t) \simeq {\kappa\over \log(E_1 t/\kappa)}
\quad.
\label{eqn:vf:relax}
\end{equation}
Free-volume relaxes logarithmically with time:
in the absence of any forcing, the system ages.

\subsection{Jamming}
In order to isolate the STZ mechanism for jamming,
I now assume that free-volume is constant, at some value $\chi$;
the evolution of the system is governed by 
equations~(\ref{eqn:stzdil:1}) and~(\ref{eqn:stzdil:2}) only,
and a constant stress $\sigma$ is applied.

These equations admit two stationary solutions:
a jammed state for which, $\dot\epsilon=\dot\gamma=0$, and
$$
\Delta = \tanh\left({\sigma\over\bar\mu}\right)
\quad;
$$
a steady deformation regime, for which,
$$
\Delta = {1\over\mu_0 \sigma}
\quad,
$$
and,
$$
\dot\epsilon=E_0\,\exp\left[{-{1\over\chi}}\right]\,
\left(\sinh\left({\sigma\over\bar\mu}\right)-{1\over\mu_0 \sigma}\,\cosh\left({\sigma\over\bar\mu}\right)\right)
\quad.
$$
The latter solution is unstable at small stresses, in which case, the material jams.
For large stresses, the jammed state loose stability, and steady deformation results.
The two solutions exchange stability at the yield stress $\sigma_y$,
which is the solution of:
$$
\tanh\left({\sigma_y\over\bar\mu}\right) = {1\over\mu_0 \sigma_y}
\quad.
$$

\subsection{Various limits}

In the following, I will consider several limiting cases of the general constitutive
equations~(\ref{eqn:stzdil:1}-\ref{eqn:stzdil:3}).

In particular, the exponential dependency of $\ddpl$ as a function of $\sigma$
will not be studied, although this non-linearity might be 
important in some cases.~\cite{saulnier02,bureau02}
Activation factors involving $\sigma$ will thus be linearized;
this is valid under the assumption that $\sigma<<\bar\mu$;
$\bar\mu$ is then incorporated into constants by taking it to unity.
It leads to:
\begin{eqnarray}
\label{eqn:stzdil:lin:1}
\ddpl&=&E_0\,\exp\left[{-{1/\chi}}\right]\,\left(\sigma-\Delta\right)\\
\label{eqn:stzdil:lin:2}
\dot\Delta&=&{\ddpl\over\epsilon_0}\,\left(1- \mu_0\,\sigma\,\Delta\right)\\
\label{eqn:stzdil:lin:3}
\dot \chi &=& -E_1 \exp\left[-{\kappa/\chi}\right]+\alpha\,\sigma\ddpl
\quad.
\end{eqnarray}

I will also consider an {\em isotropic (liquid) limit}, 
where the structural anisotropy, measured by $\Delta$ plays no role. 
This is the case either at timescales when the dynamics of $\Delta$ is too slow,
$1/\epsilon_0\to0$ (and $\Delta=0$) or because the typical stress $1/\mu_0$ vanishes. 
In this case, the rheology is determined by a single state variable:
\begin{eqnarray}
\label{eqn:stzdil:iso:1}
\ddpl&=&E_0\,\exp\left[{-{1/\chi}}\right]\,\sigma\\
\label{eqn:stzdil:iso:2}
\dot \chi &=& -E_1 \exp\left[-{\kappa/\chi}\right]+\alpha\,\sigma\ddpl
\quad.
\end{eqnarray}
The sets of equations~(\ref{eqn:stzdil:lin:1}-\ref{eqn:stzdil:lin:3})
and~(\ref{eqn:stzdil:iso:1}-\ref{eqn:stzdil:iso:2})
will be helpful to emphasize the roles of $\Delta$ versus $\chi$. 
Many aspects of glassiness will be shown to be captured by
equations~(\ref{eqn:stzdil:iso:1}) and~(\ref{eqn:stzdil:iso:2}).
Jamming, however, is not captured at this level of simplification,
and requires the existence of the variable $\Delta$.

\subsection{Experimental protocol}

A word about the preparation of the sample: 
at time $t=0$, the system is quenched from a highly 
mixed state, in which, $\Delta=0$ and $\chi$ is large.
The sample then ages during a waiting time, $t_w$ without any forcing.
After this relaxation, some test is performed on the system.

Two main experimental procedures consist in forcing the stress,
or forcing the deformation. 
If a stress, $\sigma$, is applied, the plastic deformation $\gamma(t)$ 
is determined by constitutive equations, 
and the total deformation  is, $\epsilon = \gamma + \sigma/\mu$.
If the deformation, $\epsilon(t)$, is forced, the evolution 
of $\sigma$ is given by equation~(\ref{eqn:sigma:0}), 
and couples to constitutive equations.

If a small perturbation is applied to the sample,
the response of the system results from first order terms in the 
linearized constitutive equations~(\ref{eqn:stzdil:lin:1}-\ref{eqn:stzdil:lin:3}),
or~(\ref{eqn:stzdil:iso:1}-\ref{eqn:stzdil:iso:2}).
In relaxation processes, however, the linear approximation 
is not always valid at long times,~\cite{yoshino98,derec01} 
because, the activation factors can become small, 
possibly smaller than non-linear terms.
In the current theory, all non-linear terms involve the work of plastic
deformation $\sigma\dot\gamma$. Therefore, the linear analysis is valid
so long as this work is small compared to relaxation processes.

\section{Stress relaxation after a strain increment}
Stress relaxation is monitored after a small strain step.
The step can be understood as a ramp in the deformation, which occurs
on a very small time interval around $t_w$.
The strain $\epsilon$ determines the initial value of 
the stress, $\sigma(t_w^+)=\mu\,\epsilon$.
During the relaxation, $\dot\epsilon=0$;
the subsequent evolution of the stress is governed by $\dot\sigma=-\mu\dot\gamma$, 
and accounts for the plastic adaptation of the material to the imposed strain.

The relaxation modulus, $G(t)=\sigma(t)/\sigma(t_w)$ is the correlator associated 
with rearrangement processes: it measures the correlation between the 
initial state of the material, and its state at any following time, $t$.
Assuming that the diffusion of molecules is dominated by collective rearrangements,
stress relaxation can then be compared with the decay of 
self-intermediate scattering functions, which measure the spatial 
decorrelation resulting from the diffusion of particles in the dense medium.~\cite{boon91}
The long-time behavior of correlation functions is one of the major
properties of glassy materials.~\cite{goetze92}
(See also recent reviews by Debenedetti and Stillinger,~\cite{debenedetti01}
or Angell {\it et al}.~\cite{angell00}).


\subsection{Isotropic limit}

In the isotropic limit, the constitutive equations 
are~(\ref{eqn:stzdil:iso:1}) and~(\ref{eqn:stzdil:iso:2}).
Stress relaxation is governed by the linearized equations:
\begin{eqnarray*}
\dot\sigma &=& -\mu\,E_0\,\exp\left[-{1\over\chi}\right]\sigma\\
\dot\chi &=& -E_1 \exp\left[-{\kappa\over\chi}\right]
\quad.
\end{eqnarray*}
To determine the relaxation modulus, $G=\sigma/\sigma(t_w)$,
the only relevant initial condition is the initial value of $\chi$,
when the strain is applied; this value is also determined by the initial free-volume
$\chi_0$ at time $t=0$ and by the waiting time $t_w$.
Moreover, either $E_1$ or $\mu\,E_0$ can be incorporated in the time scale,
therefore, only two effective parameters determine all the possible outcomes
of those equations: $\kappa$, and $\mu\,E_0/E_1$. While $\kappa$
measures the relative heights of entropy barriers, $\mu\,E_0/E_1$ measures 
the relative update frequencies of shear and compaction processes.

\subsubsection{Long time relaxation}

At long time, the relaxation of $\chi$ is given by~(\ref{eqn:vf:relax}),
and stress relaxation verifies:
$$
\dot\sigma = -\mu\,E_0\,\exp\left[-{1\over\chi}\right]\sigma
\simeq -\mu\,E_0\, \left({E_1\,t\over\kappa}\right)^{-1/\kappa}\sigma
\quad.
$$

The response modulus $G(t) = \sigma(t)/\sigma(t_w)$ is,
$$
G(t) \simeq \exp\left[{A\,\left(t_w^\beta-t^\beta\right)}\right]
$$
with an exponent $\beta$ which is directly related
to the ratio $\kappa$ between the heights of entropy barriers:
$$
\beta = 1-{1\over\kappa}\quad {\rm and}\quad A={\mu\, E_0\over\beta}
\,\left({E_1\over\kappa}\right)^{-1/\kappa}
\quad.
$$

The validity of linear approximation is checked by considering the time-dependency
of the work of plastic deformations $\sigma\dot\gamma$;
this term should be compared with the first term is free-volume relaxation,
$$
E_1\,\exp\left[-\kappa/\chi\right]\propto {\kappa\over t}
\quad.
$$
The stress relaxation, leads to the following evolution for the work of plastic 
deformations:
$$
\sigma\dot\gamma \propto t^{-{1/\kappa}}\;\exp\left[-2At^\beta\right]
\quad.
$$
If $\kappa>1$, or $\beta>0$, the relaxation of $\sigma\dot\gamma$ is dominated
by the exponential factor;
if $\kappa<1$, the stress goes to a constant, but $t^{-{1/\kappa}}$ decays
faster than $1/t$.
In all cases, the non-linear term decays faster than the linear term,
which justifies the linear approximation at all times.

For $\beta>0$, {\it i.e.} $\kappa>1$, the stress undergoes KWW relaxation. 
This property results directly from the existence of two 
types of saddle points of unequal heights,
when hopping motion is controlled by a logarithmically relaxing free-volume.
Measured values of exponent $\beta$ in glassy materials, 
ranging from $0.2$ to $0.5$~\cite{larson99,angell00}
are consistent with, $\kappa=v_1/v_0\in[1.25,2]$.
The ratio $\kappa$ is related to the shape of molecules,
and to the details of their interactions.
There is no {\it a priori} reason, from a phase space point of view,
to assume a particular value for $\kappa$. In this work I will systematically
study both cases in order to draw a complete picture of the phenomenology 
described by the proposed constitutive equations.

If $\kappa<1$, then $\beta<0$, and
the previous calculation shows that the stress assumes a non-vanishing 
asymptotic value after relaxation. 
In this case, $A<0$, and 
the stress relaxation towards its ultimate constant value can 
be further estimated as
a power law:
\begin{eqnarray*}
G(t)
&\simeq& \exp\left[ |A|\,\left({t^{-|\beta|}}-{t_w^{-|\beta|}}
\right)\right]\\
&\simeq& \exp\left[-|A|\,t_w^{-|\beta|}\right]\;\left(1+|A|\,t^{-|\beta|}\right)
\quad.
\end{eqnarray*}
Interpreting $G$ as a correlation function, it also means that the material 
never completely decorrelates from its initial state. 
This non-ergodicity is an indication of the glass transition as understood
in mode-coupling theory~\cite{goetze92};
it can be characterized by the asymptotic value $G(t\to\infty)$, 
the Edwards-Anderson parameter:
$$
G(t\to\infty) \simeq \exp\left[{-|A|\,t_w^{-|\beta|}}\right]
\quad,
$$
which increases with the age of the sample. 

Such a dependency of the Edwards-Anderson parameter
of the age of the sample has been observed experimentally by Bonn and coworkers
on a colloidal glass at very low concentrations.~\cite{bonn99}

\subsubsection{Short time response}

At short time, if free-volume can be assumed large or constant,
the exponential term $\exp\left[-{1/\chi}\right]$ does not vary very much,
and stress undergoes an exponential relaxation:
$$
G(t) = 
\exp\left[-\mu\,E_0\,\exp\left[-{1\over\chi_0}\right]\;\left(t-t_w\right)\right]
\quad.
$$
This relaxation occurs on a time scale of order, 
$$
\tau_\sigma = {1\over\mu\,E_0}\,\exp\left[{1\over\chi_0}\right]
\quad,
$$
which should be compared with the time scale of free-volume relaxation, $\tau_\chi$.
If $\tau_\sigma>>\tau_\chi$, free-volume relaxes faster than stress, 
and the long time, free-volume controlled, stress relaxation can be observed.
If $\tau_\sigma<<\tau_\chi$, the stress relaxes exponentially before free-volume
achieves sufficiently low values to slow down stress relaxation.
The crossover between exponential and KWW relaxation is determined by,
$\tau_\sigma\sim\tau_\chi$ or,
$$
{1\over\mu\,E_0}\,\exp\left[{1\over\chi_0}\right] \sim 
{\chi_0\over E_1}\;\exp\left[{\kappa\over\chi_0}\right]
\quad.
$$
For a large initial free-volume, this reduces to,
$$
E_1 \sim \mu\,E_0\,\chi_0
\quad;
$$
KWW relaxation can be observed for a larger value of $E_1$, 
or for sufficiently small initial value of the free-volume, $\chi(t_w)$,
(that is for sufficiently large waiting time $t_w$).

In order to look at the short and long time response of the system, 
equations~(\ref{eqn:sigma:0}),~(\ref{eqn:stzdil:iso:1}) and~(\ref{eqn:stzdil:iso:2})
are integrated numerically with a Runge-Kutta algorithm with adaptive timestep.
The system is prepared in a dilute state, $\chi=\chi_0$, which is allowed to relax 
during time $t_w$, in the absence of any forcing ($\sigma([0,t_w])=0$).
At time $t_w$, a strain $\epsilon=10^{-3}$ is applied and produces an 
elastic response, 
$\sigma(t_w^+)=\mu\epsilon$. The dynamics of free-volume and stress are then monitored.

\begin{figure}
\narrowtext
\begin{center}
\unitlength = 0.005\textwidth
\begin{picture}(100,110)(5,0)
\put(10,5){\resizebox{90\unitlength}{!}{\includegraphics{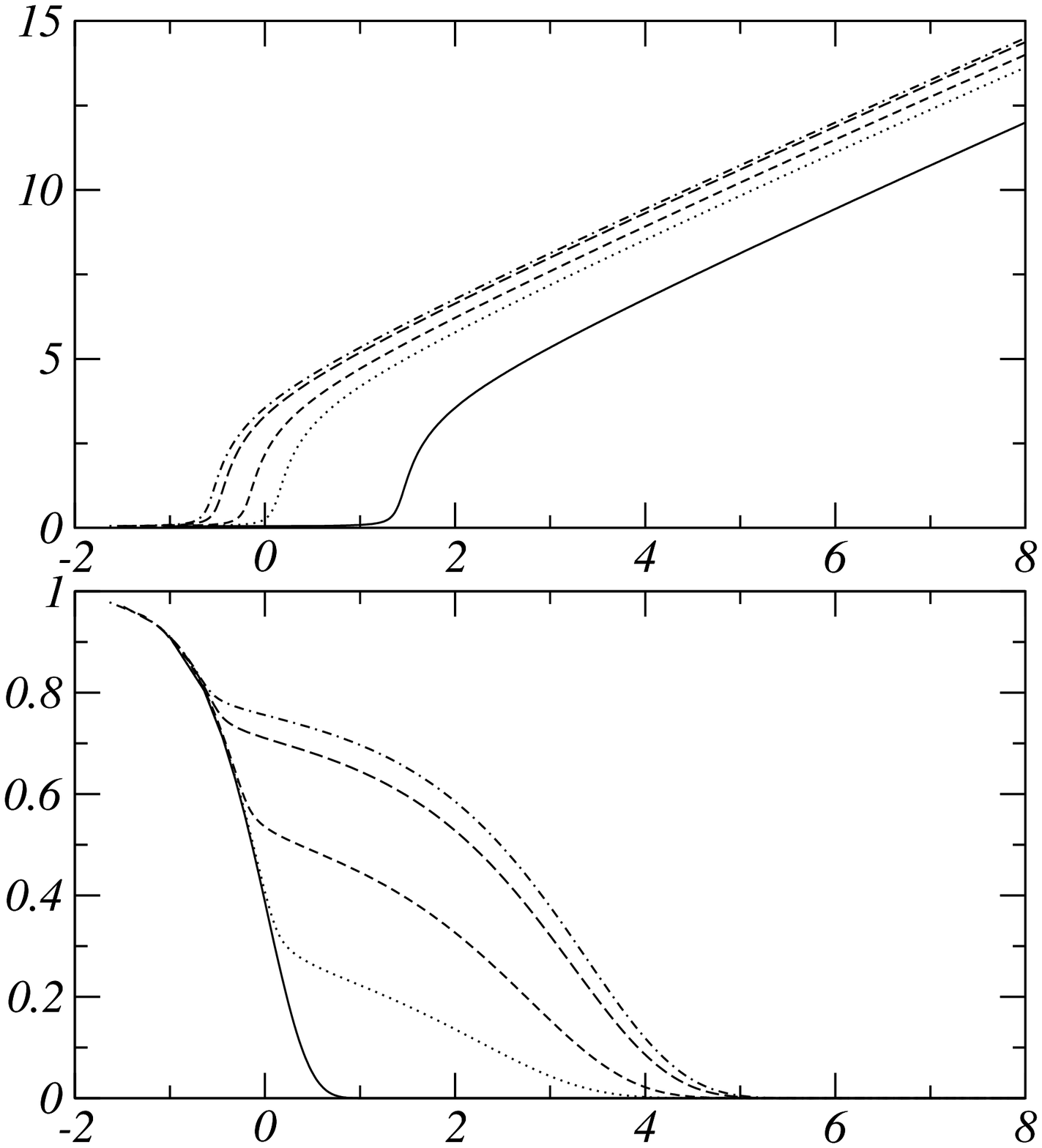}}}
\put(5,95){\makebox(0,0){\large $1/\chi$}}
\put(5,50){\makebox(0,0){\large $G$}}
\put(90,2){\makebox(0,0){\large $\log_{10}(t)$}}
\end{picture}
\end{center}
\caption{\label{fig:relax1}
Numerical integration of equations~(\ref{eqn:sigma:0}),~(\ref{eqn:stzdil:iso:1})
and~(\ref{eqn:stzdil:iso:2}) for a fixed strain $\epsilon=10^{-3}$.
Parameters are $E_0=\alpha=\mu=1$, and $\kappa=2$.
The initial stress is set to $\sigma(t_w)=\mu\epsilon=10^{-3}$, 
and $\chi_0=20$, $t_w=0$.
Top: $1/\chi$ as a function of $\log_{10}(t)$
for $E_1 = 1, 20, 40, 80, 100$ from right to left.
Bottom: relaxation spectra with $E_1$ increasing from left to right.
}
\end{figure}

\paragraph{Crossover stress.}
In a first series of tests, $t_w$ is set to 0, 
and the relaxation spectra are displayed for different values of the parameter $E_1$.
Parameters $E_0$, $\mu$ and $\alpha$ are set to 1; $\kappa=2$ and $\chi_0=20$.
The results are displayed figure~\ref{fig:relax1}.
For the first value $E_1=1$, only exponential relaxation is observed
because all the stress has been released before $\chi$ entered the logarithmic
relaxation.
The value $E_1=20$ corresponds to the crossover $\mu\,E_0\,\chi_0$:
KWW relaxation is observed at later stages of the relaxation,
with a non-vanishing values of $G$.
For a given $\chi_0$,
the stress release during the exponential part of the relaxation is
proportional to $\mu\,E_0$, hence inversely proportional to $E_1$.
The plateau in the relaxation spectrum can be observed only 
for sufficiently large values of $E_1/(\mu\,E_0)$ so that $G$ is still observable 
by the time $\chi$ enters the logarithmic relaxation.

These relaxation curves are very similar to those observed in 
MD simulations,~\cite{kob95,angell00,debenedetti01} 
or in experiments;~\cite{megen94}
the general appearance of those curves depends sensibly
on the parameters considered. 
From a practical point of view, it means that it depends
on the time window that is available, numerically or experimentally.

\begin{figure}
\narrowtext
\begin{center}
\unitlength = 0.005\textwidth
\begin{picture}(100,110)(5,0)
\put(10,5){\resizebox{90\unitlength}{!}{\includegraphics{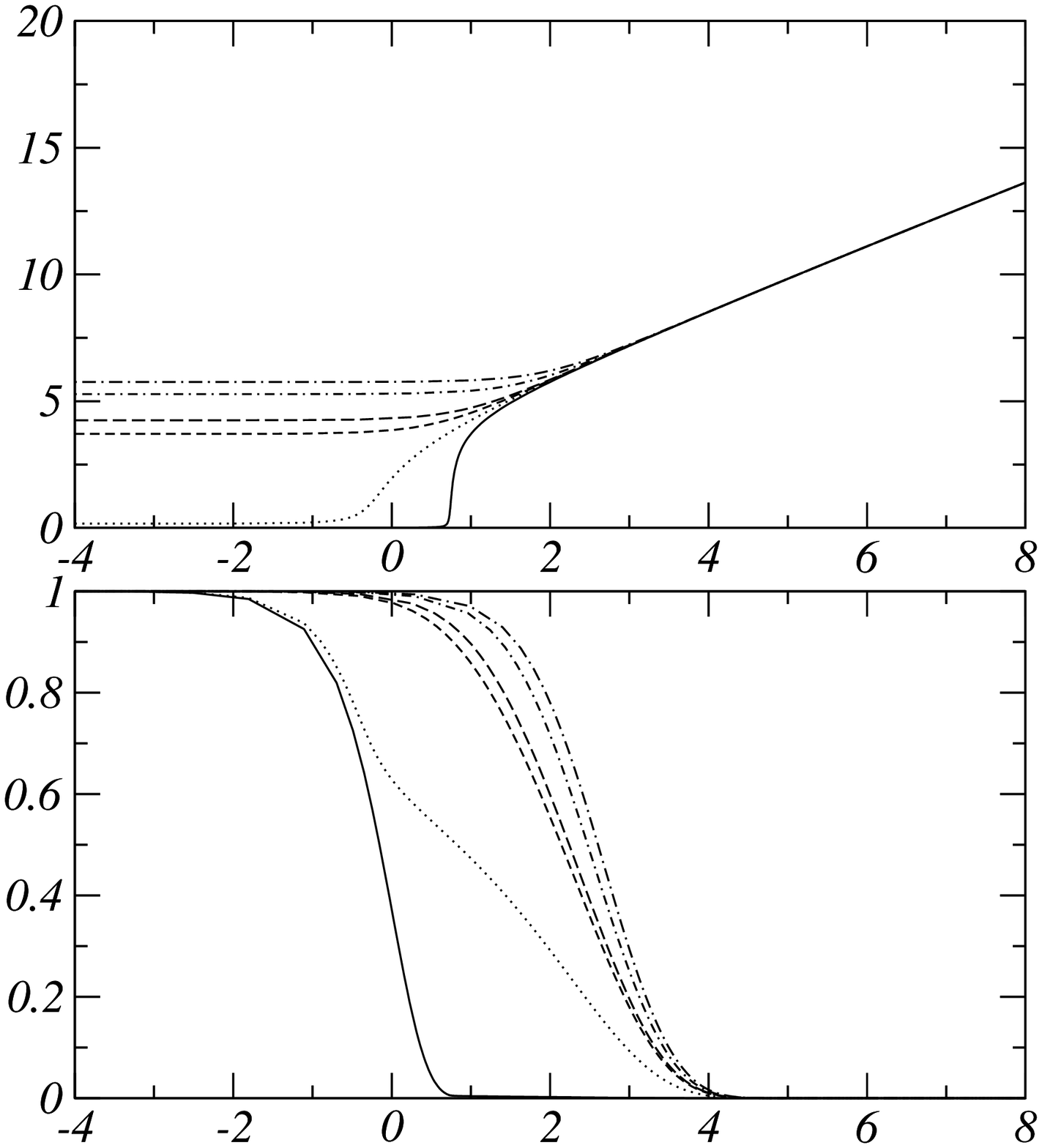}}}
\put(5,95){\makebox(0,0){\large $1/\chi$}}
\put(5,50){\makebox(0,0){\large $G$}}
\put(90,2){\makebox(0,0){\large $\log_{10}(t-t_w)$}}
\end{picture}
\end{center}
\caption{\label{fig:relax2}
Numerical integration of equations~(\ref{eqn:sigma:0}),~(\ref{eqn:stzdil:iso:1}) 
and~(\ref{eqn:stzdil:iso:2}) for a fixed strain $\epsilon=10^{-3}$.
Parameters are $E_0=\alpha=\mu=1$, $E_1=20$, and $\kappa=2$.
Initial conditions are obtained by a quench at time $t=0$ from a dilute state
($\sigma(0)=0$ and $\chi(0)=100$) followed by density relaxation without forcing during 
time $t_w$; at time $t_w$ the stress is set to $\sigma(t_w)=\mu\epsilon=10^{-3}$.
Top: $1/\chi$ as a function of $\log_{10}(t-t_w)$
with $t_w = 0, 5, 10, 15, 50, 100$ from bottom to top.
Bottom: relaxation spectra for increasing $t_w$ from left to right.
}
\end{figure}

\paragraph{Aging}
The crossover between exponential relaxation can also be studied by varying
the initial free-volume. This amounts to varying the waiting time $t_w$
for a given $\chi(t=0)$.
In order to observe a plateau in the relaxation spectrum, parameter $E_1$ is set to 20,
and parameters $E_0$, $\mu$ and $\alpha$ are set to 1; $\kappa=2$.
In order to obtain an exponential relaxation for short waiting times, 
and to observe a wide variety of behavior, the initial value of $\chi$ is set to 100.
The results are displayed figure~\ref{fig:relax2}. With this choice of parameter,
the curves resemble experimental data.~\cite{angell00,gopal99,hebraud97,megen98,cipelletti00} 

\begin{figure}
\narrowtext
\begin{center}
\unitlength = 0.005\textwidth
\begin{picture}(100,110)(5,0)
\put(10,5){\resizebox{90\unitlength}{!}{\includegraphics{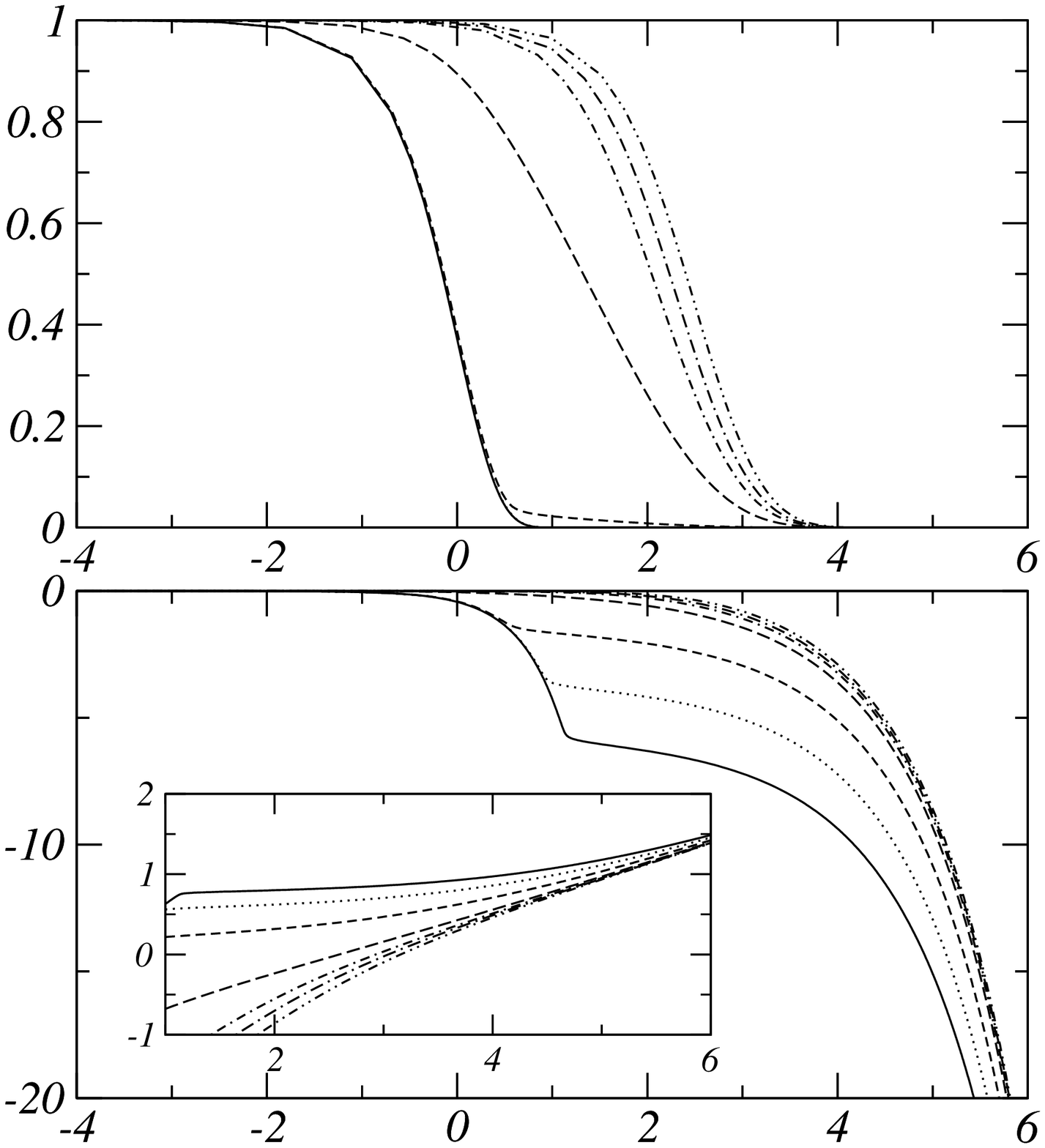}}}
\put(5,95){\makebox(0,0){\large $G$}}
\put(5,50){\makebox(0,0){\large $\log_{10}(G)$}}
\put(90,2){\makebox(0,0){\large $\log_{10}(t-t_w)$}}
\end{picture}
\end{center}
\caption{\label{fig:relax}
Numerical integration of equations~(\ref{eqn:sigma:0}),~(\ref{eqn:stzdil:iso:1}) 
and~(\ref{eqn:stzdil:iso:2}) for a fixed strain $\epsilon=10^{-3}$.
Parameters are $E_0=E_1=\alpha=\mu=1$, and $\kappa=1.5$.
Initial conditions are obtained by a quench at time $t=0$ from a dilute state
($\sigma(0)=0$ and $\chi(0)=10$) followed by density relaxation without forcing during 
time $t_w$; at time $t_w$ the stress is set to $\sigma(t_w)=\mu\epsilon=10^{-3}$.
Top: $1/\chi$ as a function of $\log_{10}(t-t_w)$
with $t_w = 0, 5, 10, 50, 100$ from bottom to top.
Bottom: relaxation spectra for increasing $t_w$ from bottom to top.
There is no inconsistency in the large logarithmic scale for $\log_{10}(G)$: 
the scale is determined up to the reduced parameter $A$.
Inset: $\log_{10}(-\log_{10}(G))$ as a function of $\log_{10}(t-t_w)$,
with $t_w$ increasing from top to bottom.
The spectra with the longest waiting time are already in the KWW regime.
}
\end{figure}

Such curves strongly depend on the available experimental window, and
on the width of the crossover region which is unobservable in a log-lin plot for 
smaller values of $E_1/(\mu\,E_0)$. 
To emphasize this point,
another set of relaxation spectra is presented figure~\ref{fig:relax}
for values of the parameters: $E_0=\alpha=\mu=1$, $E_1=8$, and $\kappa=2$.
The initial condition is $\chi(0)=100$ and waiting times are:
$t_w = 0, 5, 10, 15, 50, 100$. The relaxation spectra are presented in log-lin,
log-log, and log-log(log) plot; the crossover between exponential and KWW relaxation 
is observable around values of the relaxation modulus 
of order $10^{-5}$, which is featureless in a log-lin plot.

\begin{figure}
\narrowtext
\begin{center}
\unitlength = 0.005\textwidth
\begin{picture}(100,110)(5,0)
\put(10,5){\resizebox{90\unitlength}{!}{\includegraphics{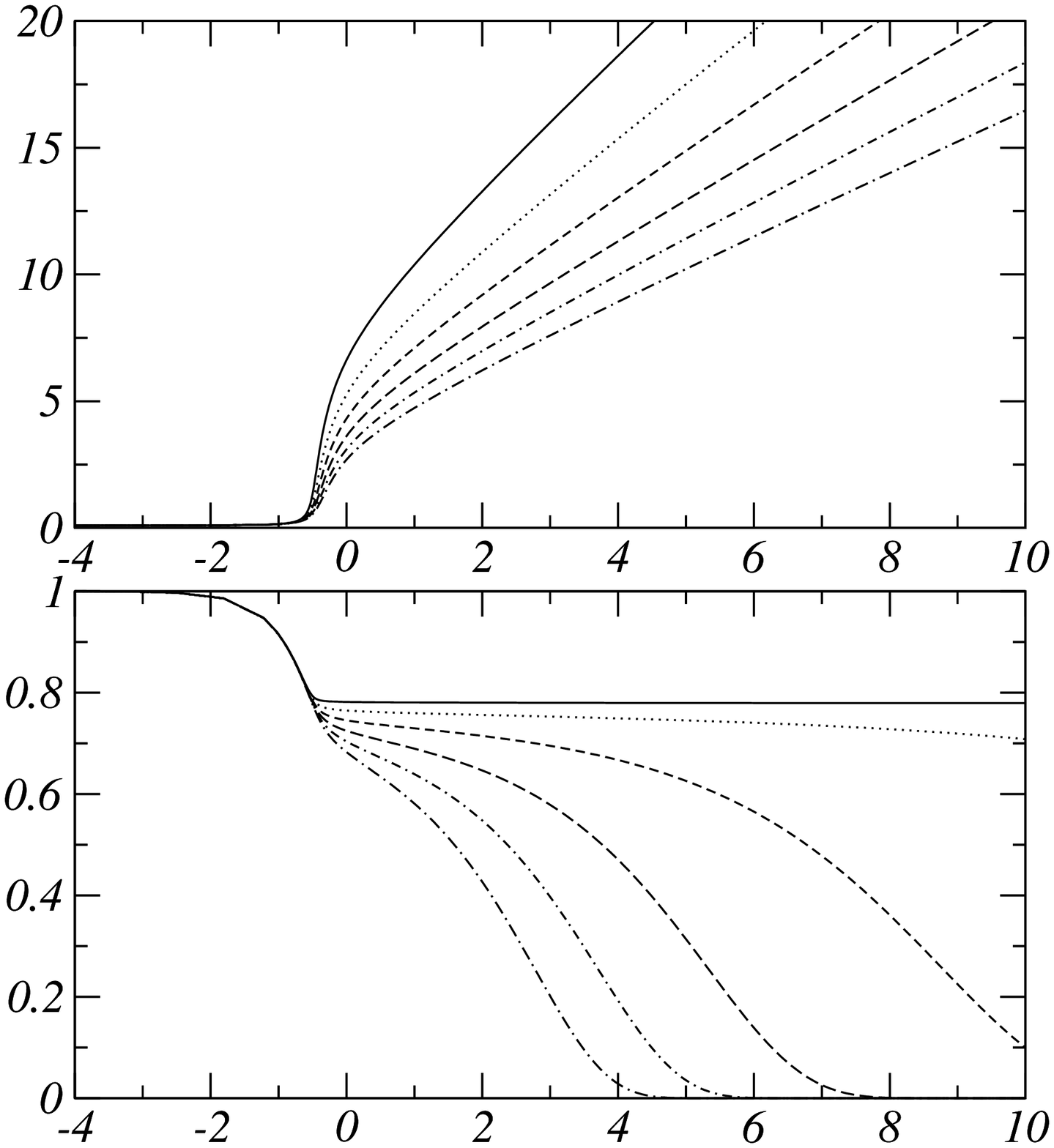}}}
\put(5,95){\makebox(0,0){\large $1/\chi$}}
\put(5,50){\makebox(0,0){\large $G$}}
\put(90,2){\makebox(0,0){\large $\log_{10}(t-t_w)$}}
\end{picture}
\end{center}
\caption{\label{fig:relax3}
Numerical integration of equations~(\ref{eqn:sigma:0}),~(\ref{eqn:stzdil:iso:1}) 
and~(\ref{eqn:stzdil:iso:2}) for a fixed strain $\epsilon=10^{-3}$.
Parameters are $E_0=\alpha=\mu=1$ and $E_1=10$; $\kappa$ varies.
Initial conditions are obtained by a quench at time $t=0$ from a dilute state
($\sigma(0)=0$ and $\chi(0)=100$) followed by density relaxation without 
forcing during time $t_w$; 
at time $t_w$ the stress is set to $\sigma(t_w)=\mu\epsilon=10^{-3}$.
Top: $1/\chi$ as a function of $\log_{10}(t-t_w)$
with $\kappa = 1, 1.2, 1.4, 1.6, 1.8, 2$ from top to bottom.
Bottom: relaxation spectra for increasing $\kappa$ from top to bottom.
}
\end{figure}

\paragraph{Ergodic to non-ergodic transition.}
I now consider the dependency of relaxation spectra on the parameter $\kappa$.
This parameter is of primary importance in the current approach and
is directly related to the existence of entropy barriers of unequal heights 
in the phase space.
The parameters are taken to be, $E_0=\alpha=\mu=1$, $E_1=10$, 
the initial condition is $\chi(0)=100$, and the waiting
is set to $t_w=10$. This choice is primarily motivated to obtain easily
observable features on the relaxation spectrum.
The results are displayed figure~\ref{fig:relax3}.
When $\kappa$ goes to 1, the system presents a transition to non-ergodic behavior.
For $\kappa<1$, the asymptotic value of the stress is no longer vanishing.
At the point $\kappa=1$, the long time relaxation of $\sigma$ verifies:
$$
\dot\sigma = -\mu\,E_0\,\exp\left[-{1\over\chi}\right]\sigma
\simeq -\mu\,{E_0\over E_1\,t}\sigma
\quad,
$$
whence an asymptotic power law relaxation:
$$
G(t) \simeq \left({t\over t_w}\right)^{-{\mu\,E_0/ E_1}}
\quad.
$$

\subsection{General case}

KWW relaxation is also achieved with the complete set of 
equations~(\ref{eqn:stzdil:lin:1}-\ref{eqn:stzdil:lin:3}).
At time $t=0$, $\Delta(0)=0$, and no constraint is applied until time $t_w$:
$\sigma([0,t_w]) = 0$.
During the ramp, the deformation of the material is essentially elastic, 
no rearrangement occurs, hence, $\Delta(t_w^+)=0$, while the imposed 
strain determines the initial value the stress: $\sigma(t_w^+)=\mu\epsilon$.

During the ensuing relaxation, $\dot\sigma=-\mu\dot\gamma$ and,
for a small deformation, $\dot\Delta\simeq\dot\gamma/\epsilon_0$. 
It results that, at all times, the quantity $\sigma+\mu\,\epsilon_0\,\Delta$ 
is constant:
$$
\sigma+\mu\,\epsilon_0\,\Delta = \epsilon\, U_0
\quad.
$$
Here, $U_0=\mu$,
but this expression may vary, depending on the preparation of a sample.
To treat initial values of $\sigma$ and $\Delta$ in the greatest generality,
the initial value for the difference $\sigma-\Delta$ is denoted $\epsilon\,S_0$
(and here, $S_0=\mu$).

At long times, the dynamics of $\Delta$ and $\sigma$ verify,
\begin{eqnarray*}
\dot\sigma-\dot\Delta &\simeq& -\left(\mu+{1\over\epsilon_0}\right)\,E_0\,
\exp\left[-{1\over\chi}\right]\left(\sigma-\Delta\right)\\
&\simeq& -\left(\mu+{1\over\epsilon_0}\right) 
\,E_0\,\left({E_1\,t\over\kappa}\right)^{-1/\kappa}\left(\sigma-\Delta\right)
\end{eqnarray*}
whence,
$$
\sigma-\Delta \simeq \epsilon\,S_0\,\exp\left[{A'\,(t_w^\beta-t^\beta)}\right]
$$
with,
$$
\beta = 1-{1\over\kappa}\quad {\rm and}\quad 
A'=\left(\mu+{1\over\epsilon_0}\right)\,{E_0\over\beta}\,
\left({E_1\over\kappa}\right)^{-1/\kappa}
\quad.
$$
The evolution in time of the quantity $\sigma-\Delta$ 
is governed by the same equation as the dynamics of $\sigma$ in the isotropic limit.
Therefore, all the previous discussion presented in the isotropic limit pertains
to the general case, up to a constant term, which is fixed by the conservation
of the quantity $\sigma+\mu\,\epsilon_0\,\Delta$.
The dynamics of $\Delta$ and $\sigma$ follow.

In particular, the evolution of $\Delta$ reads,
$$
\Delta(t) \propto \epsilon\;{U_0-S_0\exp\left[{A'\,(t_w^\beta-t^\beta)}\right]\over 1+\mu\epsilon_0}
\quad,
$$
and the relaxation modulus verifies,
$$
G(t) \propto {U_0+\mu\epsilon_0\,S_0\,\exp\left[{A'\,(t_w^\beta-t^\beta)}\right]\over 1+\mu\epsilon_0}
\quad.
$$
The isotropic expression for $G(t)$ is recovered by taking $\epsilon_0\to\infty$.
Note that the relaxation of the stress could lead to a negative value if $U_0<0$,
which depends on the preparation of the sample.

For $\beta>0$ ({\it i.e.} $\kappa>1$), the stress undergoes 
a KWW relaxation towards a non-vanishing value, which
depends on the properties of the material and on the preparation of the sample.
For $\beta<0$ power law relaxation is observed towards a non-vanishing value
of the stress.

In the current work, two origins are therefore identified for 
the existence of a non-vanishing stress at long times: 
the creation of mechanical anisotropy of the contact network (jamming), 
or the existence of a ergodic to non-ergodic transition
for the hopping motion in phase space (freezing). 
Recent observations by Bonn {\it et al} indicate that for a given material, 
jamming could not be identified only to a loss of ergodicity, 
and that some amount of jamming could coexist 
with liquid-like behavior.~\cite{bonn02a,bonn02b} 
Such features are allowed in the current theory and show the importance 
of the STZ mechanism for soft glassy materials.

The validity of the linear approximation is checked by considering the 
long times relaxation of the rate of plastic deformation:
$$
\dot\gamma \propto t^{-{1/\kappa}}\;\exp\left[{-A'\,t^\beta}\right]
$$
and the work of plastic deformation decays like $\dot\gamma$ since $\sigma$ goes to a constant;
the non-linear term in equation~(\ref{eqn:stzdil:lin:3}) 
is negligible at all times.

\section{Response to a constant strain rate}

\subsection{\label{sec:steady} Steady plastic flow}
When a constant strain rate $\dot\epsilon$ is applied,
the material is driven strongly out-of-equilibrium;
the response of the system involves non-linear terms.
I study here the stationary states of a sheared material, in which, $\dot\gamma=\dot\epsilon$.
In the first sections, I make a very primitive stability analysis of those stationary
states, by assuming that $\dot\gamma=\dot\epsilon$ is fixed at all times, and that the
stress is given by the constitutive relation~(\ref{eqn:stzdil:lin:1}) 
or~(\ref{eqn:stzdil:iso:1}); this allows to decouple equation~(\ref{eqn:sigma:0}).

The existence of an unstability in the complete set of equations accounts 
for the emergence of stick-slip motion in boundary lubrication.~\cite{lemaitre01a}
This question will be briefly discussed in section~\ref{sec:hopf},
but raises so many issues that it will be tackled thoroughly in another publication;
for the sake of completeness, 
the calculation of the Hopf criterion is given in appendix~\ref{app:hopf}.

\subsubsection{Isotropic limit}
In the limit $\mu_0\to\infty$ (no yield stress), the relation between
shear rate and shear stress reads (from~(\ref{eqn:stzdil:iso:1})):
$$
\sigma = {\dot\epsilon \over E_0}\, \exp\left[{1\over\chi}\right]
\quad.
$$
Eliminating $\sigma$ between equations~(\ref{eqn:stzdil:iso:1}) and~(\ref{eqn:stzdil:iso:2}), 
yields:
\begin{equation}
\dot\chi = -E_1 \exp\left[{-{\kappa\over\chi}}\right] + \alpha {\dot\epsilon^2\over E_0}\, \exp\left[{1\over\chi}\right]
\label{eqn:chidyn}
\end{equation}
which is positive iff,
$$
{\alpha\dot\epsilon^2\over E_0\,E_1} > e^{-{\kappa+1\over\chi}}
\quad.
$$
In the stationary flow, the relation between shear strain and volume reads,
$$
\dot\epsilon = \sqrt{E_0\,E_1\over\alpha}
\,\exp\left[{-{\kappa+1\over2\chi}}\right]
\quad.
$$
For future use, I introduce the critical rate of deformation,
$$
\dot\epsilon^* = \sqrt{E_0\,E_1\over\alpha}
\quad.
$$
\begin{figure}
\narrowtext
\begin{center}
\unitlength=0.005\textwidth
\begin{picture}(110,60)(8,-8)
\put(13,40){\makebox(0,0){\large$\dot\epsilon^*$}}
\put(10,48){\makebox(0,0){\large$\dot\epsilon$}}
\put(85,-13){\makebox(0,0){\large$\chi$}}
\put(10,-13){\resizebox{90\unitlength}{!}{\includegraphics{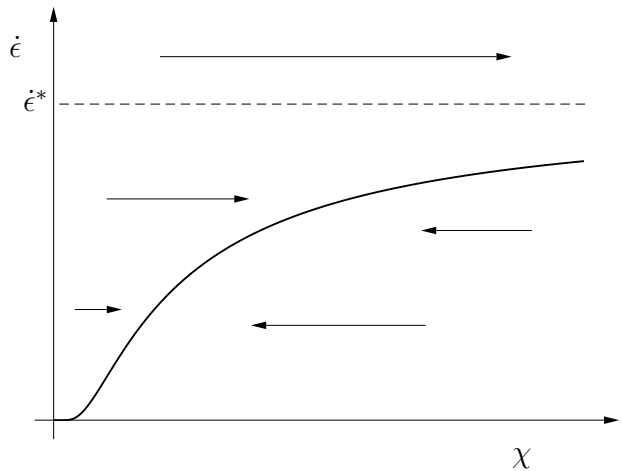}}}
\end{picture}
\end{center}
\caption{\label{fig:vf}
Bifurcation diagram for equation~(\protect\ref{eqn:chidyn}).
$\dot\epsilon$ is drawn as a function of $\chi$,
and the strain rate $\dot\epsilon$ is fixed.
For $\dot\epsilon> \dot\epsilon^*$,
no solution exists, $\dot\chi>0$, the system blows up;
for $\dot\epsilon < \dot\epsilon^*$ the system admits one stable solution.}
\end{figure}
The bifurcation diagram for $\chi$ is presented figure~\ref{fig:vf}.
If $\dot\epsilon> \dot\epsilon^*$, then $\dot\chi>0$ at all times: 
free-volume diverges, the system blows up.
If $\dot\epsilon < \dot\epsilon^*$, the equation for $\chi$ admits one solution:
$$
\chi = -{\kappa+1\over 2\,\left(\log{\dot\epsilon}-\log{\dot\epsilon^*}\right)}
\quad.
$$
The existence of a stationary value for the free-volume indicates 
that for any finite rate of deformation, aging stops.
In fact, this is not completely exact if a vanishingly small shear rate
was applied, such that the associated steady free-volume is so small, 
that it is not accessible at reasonable timescales. 
As a consequence, for very small $\dot\gamma$,
the apparent viscosity saturates to a constant value. This is a common feature
of complex liquids (see {\it e.g.}~\cite{larson99}).

In the steady state the value of the shear stress is,
$$
\sigma= {\dot\epsilon^{n}\over E_0}\,\left({\alpha\over E_0 E_1}\right)^{{n-1\over2}}
\quad,
$$
with,
$$
n={\kappa-1\over\kappa+1}
\quad.
$$
The system behaves like a power law fluid;
the exponent $n$ of the power law viscosity is
related to the exponent $\beta$ of KWW relaxation by,
$$
n = {\beta\over 2-\beta}
\quad;
$$
those two exponents are directly related to the slope $\kappa$ of $1/\chi$
as a function of $\ln t$.

The assumptions underlying the current approach 
do not permit to evaluate the activation volumes associated with density relaxation
and shear motion. However, the single parameter $\kappa$ determines two 
exponents associated with the linear and non-linear response of a strained material.

The existence of such a relation constitutes a important result of the current theory,
although the form given to this relation depends on non-generic features:
how non-linear dilatancy is introduced in the equation of motion for free-volume; 
how the update frequencies (here taken constant) are written.
In a hard-sphere material, for example, collision frequencies
depend on free-volume as $1/\chi$, and the relation between $\sigma$ and 
$\dot\epsilon$ is expected to be modified by logarithmic corrections.

Molecular dynamics simulations of sheared fluid are consistent with power law rheologies
for $n$ equal to 0.1~\cite{berthier00,berthier01} or 0.2~\cite{yamamoto97,yamamoto98}
Recent measurement of power law rheology have also been 
obtained on a colloidal glass,
leading to values of $n$ ranging from 0.1 to 0.35.~\cite{bonn02a,bonn02b}
A value of $\beta$ of order $0.4$ corresponds to $n=0.25$, which seems 
in reasonable agreement with the currently available data.

The theory of structural rearrangement described here, 
is expected to present some degree of universality, 
and hold, in a given form,  for wide classes of materials.
What this model shows is that there is an intimate relation between 
time-logarithmic density relaxation, KWW relaxation,
and power law rheology of a dense material, and that this relation may depend
only on generic features for a given class of materials.
Therefore the existence of a relation between $\beta$ and $n$
could be tested by considering families of materials. 
The experimental or numerical evaluation of such a relation 
appears as an important mean of characterization of the coupling between 
two major modes of deformation.

\subsubsection{General case}

The constitutive equations are now~(\ref{eqn:stzdil:lin:1}-\ref{eqn:stzdil:lin:3}).
In the steady state,
$$
\Delta = {1\over\mu_0\sigma}
\quad;
$$
plugging expression~(\ref{eqn:stzdil:lin:1}) for the plastic deformation 
in equation~(\ref{eqn:stzdil:lin:3}) leads to the following relation between $\chi$
and $\sigma$ in the steady state:
$$
\exp\left[{1-\kappa\over\chi}\right]={\alpha\,E_0\over E_1}\,\left(\sigma^2-{1\over\mu_0}\right)
\quad.
$$
Using~(\ref{eqn:stzdil:lin:3}) again, 
$\dot\epsilon$ can then be written as a function of $\sigma$:
$$
\dot\epsilon = 
{E_1\over\alpha}\;
\left(\alpha\,E_0\over E_1\right)^{{\kappa\over\kappa-1}}
\;{1\over\sigma}\;\left(\sigma^2-{1\over\mu_0}\right)^{{\kappa\over\kappa-1}}
\quad.
$$
For large $\sigma$ or $\dot\epsilon$, the system behaves as a power law fluid, 
with the same exponent $n$ as in the isotropic limit.
However, $\dot\epsilon$ vanishes for a non-zero value of $\sigma$, 
which determines the yield stress:
$$
\sigma_y = {1\over\sqrt{\mu_0}}
\quad.
$$

The following expression can also be obtained for $\dot\epsilon$ as a function of $\chi$:
$$
\dot\epsilon = {
\sqrt{E_0\,\mu_0}\,E_1\;\exp\left[-{\kappa/ 2\chi}\right] 
\over
\sqrt{\alpha}\;
\sqrt{E_0\alpha\,\exp\left[{\kappa/\chi}\right] 
		    + E_1\mu_0\,\exp\left[{1/\chi}\right]}
}
\quad,
$$
which is an increasing function of $\chi$, and which has a maximum for $\chi\to\infty$:
$$
\dot\epsilon^*=
{\sqrt{E_0\,\mu_0}\,E_1
\over
\sqrt{\alpha}\;
\sqrt{E_0\alpha+ E_1\mu_0}
}
$$
For $\dot\epsilon>\dot\epsilon^*$, like in the isotropic case,
the shear deformation leads to a constant increase of free-volume, 
and the system blows up.

\subsection{Transients}

When a shear rate is suddenly imposed on a material, 
the transient dynamics often lead to a first rise of the stress before 
a dynamical yield stress is reached at which plastic deformations begin;
once plastic deformation occur, the stress relaxes towards its strain-rate
dependent asymptotic value.~\cite{larson99}
These features are observed in sheared bulk solids~\cite{hassan95}
but also in sheared thin films,~\cite{drummond00,drummond01,persson00}
granular materials~\cite{losert00}
or in solid friction.~\cite{bureau02}
Another generic property of transient regimes is that 
the dynamical yield stress depends on the age of the material: 
during aging, a material strengthens.

These phenomena are captured by the proposed equations.
Strain softening and strengthening result essentially from the equation of motion 
for free-volume and are captured by the isotropic equations. 

\begin{figure}
\narrowtext
\begin{center}
\unitlength = 0.005\textwidth
\begin{picture}(100,110)(5,0)
\put(10,5){\resizebox{90\unitlength}{!}{\includegraphics{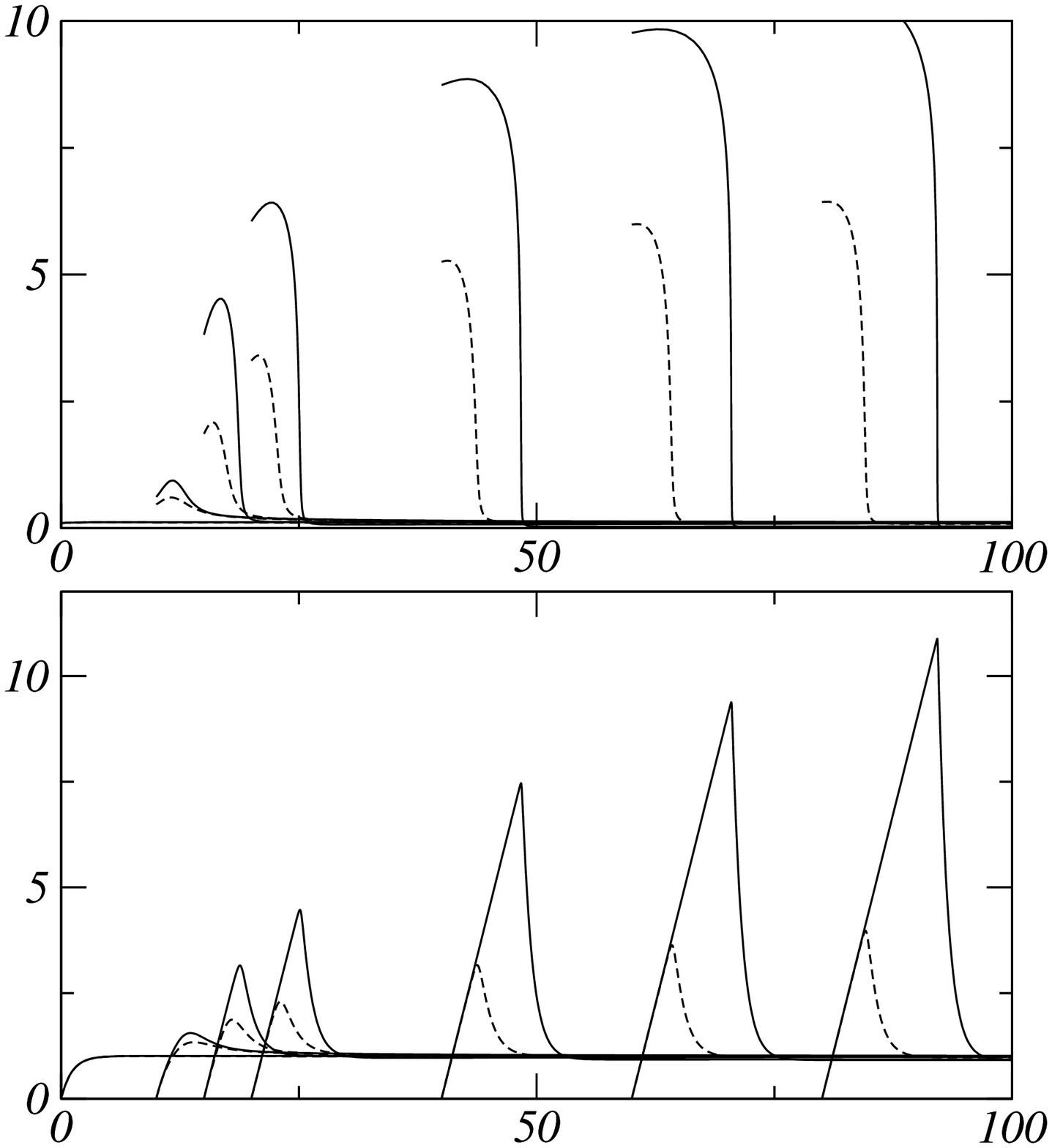}}}
\put(5,95){\makebox(0,0){\large $1/\chi$}}
\put(5,50){\makebox(0,0){\large $\sigma$}}
\put(90,2){\makebox(0,0){\large $t$}}
\end{picture}
\end{center}
\caption{\label{fig:bump}
Numerical integration of equations~(\ref{eqn:sigma:0}),~(\ref{eqn:stzdil:iso:1}) 
and~(\ref{eqn:stzdil:iso:2}) for a fixed strain rate $\dot\epsilon=0.9$.
Parameters are $E_0=E_1=\alpha=\mu=1$, and values of $\kappa$:
$\kappa=0.8$ (solid lines) and $\kappa=1.2$ (dashed lines).
No stress is applied during the initial density relaxation of the material 
from, $\chi(0)=10$.
At various times, $t_w$, the strain rate is suddenly applied and the ensuing
dynamics of $1/\chi$ (top) and  $\sigma$ (bottom) are displayed.
The strong glass $\kappa<1$ presents larger of the dynamical yield, 
which result from the smaller values of $\chi$ reached.
}
\end{figure}

Figure~\ref{fig:bump} present some examples of transient dynamics
obtained by numerical integration of
equations~(\ref{eqn:sigma:0}),~(\ref{eqn:stzdil:iso:1}-\ref{eqn:stzdil:iso:2}).
Parameters are set to $E_0=E_1=\mu=\alpha=1$.
The initial value for the free-volume (at time $t=0$) is set to $\chi(0)=10$, 
and the system relaxes without forcing during a time $t_w$.
At time $t_w$, a constant strain rate is applied.
To prevent the blow-up of the system,
$\dot\epsilon$ must be smaller than $\dot\epsilon^*$,
which equals $1$ with the chosen parameters;
for this reason, the strain rate is set to $\dot\epsilon=0.9$.
Two different values of the parameter $\kappa$ have been used to explore
the two important cases when $\kappa$ is larger or smaller than 1.
The stress is displayed as a function of the total time $t$ 
since preparation of the sample.

A dynamical yield stress clearly emerges for a sufficiently aged material, 
and increases with time. The yielding of the material is accompanied by a sudden
rise of the free-volume allowing transformation rates to display non-vanishing values.


These transients result from a rather complex interplay between the dynamics
of free-volume and shear transformation. The complete study of this process
may lead to many developments that would divert this work from the main questions
that I would like to address, and I prefer to leave the discussion here.

\subsection{\label{sec:hopf} Stick-slip instability}

The stability analysis of equations~(\ref{eqn:stzdil:lin:1}-\ref{eqn:stzdil:lin:3}) 
and~(\ref{eqn:sigma:0}) shows the existence of a Hopf bifurcation which is responsible 
for oscillatory response to a given strain rate.
This leads to some understanding of stick-slip motion in sheared 
lubricants~\cite{thompson92,yoshizawa93,demirel96a,demirel96b,drummond00,drummond01}
or in granular materials.~\cite{thompson91,nasuno97,nasuno98}
In this work, I mention this instability for the sake of completeness,
but it raises many questions, and will require a more thorough consideration.
The calculation of the Hopf criterion of bifurcation is given 
in appendix~\ref{app:hopf}.

\begin{figure}
\narrowtext
\begin{center}
\unitlength=0.005\textwidth
\begin{picture}(110,60)(8,-8)
\put(6,40){\makebox(0,0){\large$\mu_{\rm hopf}$}}
\put(68,-13){\makebox(0,0){\large$\dot\epsilon^*$}}
\put(85,-13){\makebox(0,0){\large$\dot\epsilon$}}
\put(10,-13){\resizebox{90\unitlength}{!}{\includegraphics{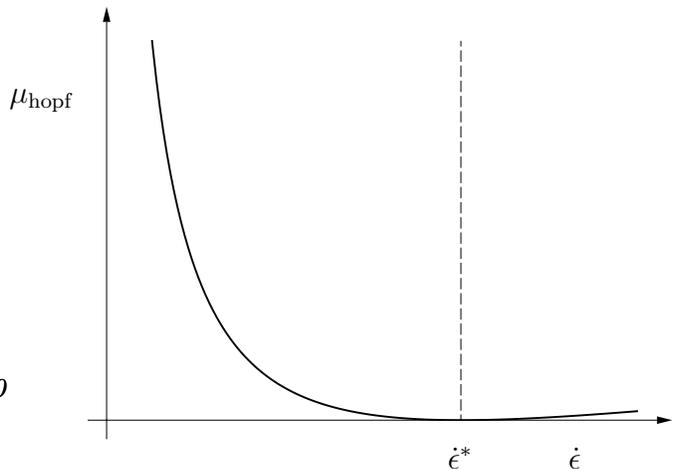}}}
\end{picture}
\end{center}
\caption{\label{fig:hopf:iso} 
Phase diagram for equations~(\ref{eqn:sigma:0}),~(\ref{eqn:stzdil:iso:1}) 
and~(\ref{eqn:stzdil:iso:2}), and $\kappa<1$.
The solid line denotes the curve $\mu_{\rm hopf}$ below 
which stick-slip motion occurs. $\mu_{\rm hopf}$ vanishes for
$\dot\epsilon=\dot\epsilon^*$ (dashed line)
where steady plastic regime disappears, leading
the system to break-up.
}
\end{figure}

In the isotropic limit, 
the steady plastic deformation is stable iff $\mu>\mu_{\rm hopf}$, with:
$$
\mu_{\rm hopf} = {E_1\over E_0}\,{1-\kappa\over(\kappa+1)^2}\;
\left({\alpha\,\dot\epsilon^2\over E_0\,E_1}\right)^{{\kappa-1\over\kappa+1}}\;
\ln\left[{\alpha\,\dot\epsilon^2\over E_0\,E_1}\right]^2
\quad.
$$
For $\kappa>1$, $\mu_{\rm hopf}<0$: there is no Hopf bifurcation.
For $\kappa<1$, $\mu_{\rm hopf}>0$ on the interval $[0,\dot\epsilon^*]$,
and the steady state becomes unstable iff $\mu<\mu_{\rm hopf}$.
This criterion is represented figure~\ref{fig:hopf:iso}.
For driving strain rates above  $\dot\epsilon^*$, the system looses stability,
because there is no stationary solution for $\chi$.
Below $\dot\epsilon^*$, the Hopf criterion separates a domain of stability,
above $\mu_{\rm hopf}$ where the system reaches steady plastic deformation,
and and domain below this curve where stick-slip motion is achieved.

\begin{figure}
\narrowtext
\begin{center}
\unitlength = 0.005\textwidth
\begin{picture}(90,90)(0,0)
\put(10,5){\resizebox{70\unitlength}{!}{\includegraphics{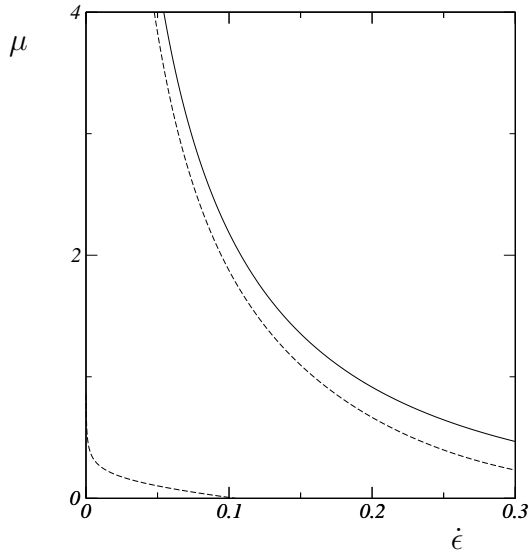}}}
\put(5,75){\makebox(0,0){\large $\mu$}}
\put(70,2){\makebox(0,0){\large $\dot\epsilon$}}
\end{picture}
\end{center}
\caption{\label{fig:hopf:lin}
Phase diagram for equations~(\ref{eqn:sigma:0}),~(\ref{eqn:stzdil:lin:1}-\ref{eqn:stzdil:lin:3}) and $\kappa<1$.
Parameters used are, $E_0=E_1=\alpha=\epsilon_0=\mu_0=1$, and $\kappa=0.8$.
The solid line denotes the Hopf criterion $\mu_{\rm hopf}$ in the isotropic
limit. The dashed lines indicates the Hopf criterion $\mu_{\rm hopf}^\pm$
with the complete set of equations.
}
\end{figure}

The introduction of the variable $\Delta$ in the set of equations~(\ref{eqn:sigma:0}),~(\ref{eqn:stzdil:lin:1}-\ref{eqn:stzdil:lin:3}) 
slightly modifies this picture (see appendix~\ref{app:hopf}).
Two limit curves $\mu_{\rm hopf}^\pm$ emerge, which delimit the unstable 
domain in the plane $(\dot\epsilon,\mu)$. 
These curves are displayed figure~\ref{fig:hopf:lin} along with the previous curves $\mu_{\rm hopf}$ of the isotropic limit.
The unstable domain is a subspace of what it is in the isotropic limit,
and the system present a creeping zone, close to $\dot\epsilon=0$,
when steady sliding is stable for very small values of the strain rate.

\begin{figure}
\narrowtext
\begin{center}
\unitlength = 0.005\textwidth
\begin{picture}(100,110)(5,0)
\put(10,5){\resizebox{90\unitlength}{!}{\includegraphics{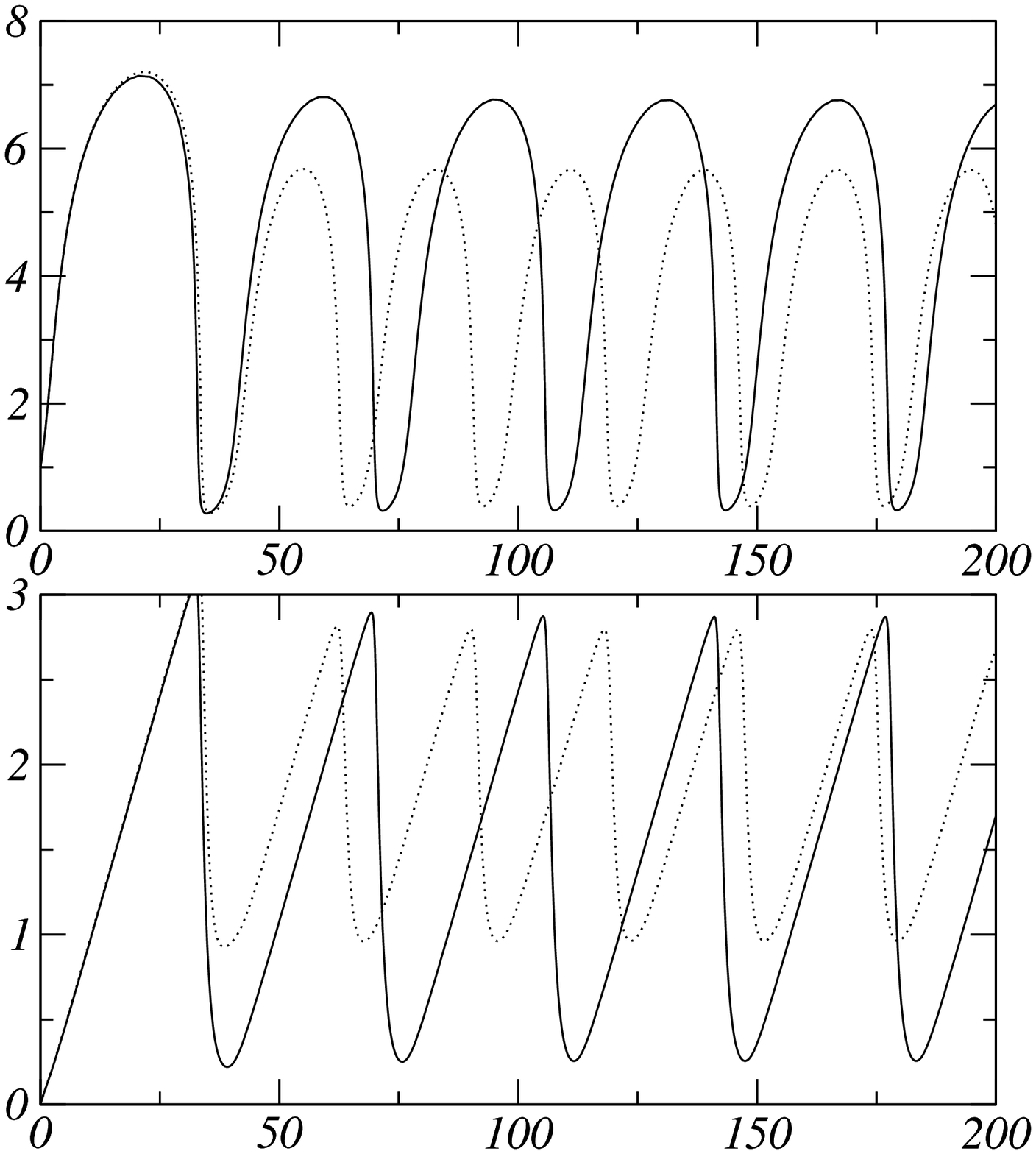}}}
\put(5,95){\makebox(0,0){\large $1/\chi$}}
\put(5,50){\makebox(0,0){\large $\sigma$}}
\put(90,2){\makebox(0,0){\large $t$}}
\end{picture}
\end{center}
\caption{ \label{fig:stickslip}
Solid lines: 
Numerical integration of equations~(\ref{eqn:sigma:0}),~(\ref{eqn:stzdil:iso:1}) 
and~(\ref{eqn:stzdil:iso:2}) for a fixed strain rate $\dot\epsilon=0.1$.
Parameters are $E_0=E_1=\alpha=\mu=1$, and $\kappa=0.8$.
Initial value of the free-volume is $\chi=10$.
The regime of steady plastic deformation is unstable and leads to stick-slip motion.
Fast relaxations of the stress result from sudden dilatancy of the material.
Dashed lines: Same for equations~(\ref{eqn:sigma:0}),~(\ref{eqn:stzdil:lin:1}-\ref{eqn:stzdil:lin:3}) with the same parameters and, $\epsilon_0=\mu_0=1$.
}
\end{figure}

Some examples of stick-slip motion are displayed figure~\ref{fig:stickslip},
with the set of parameters used in the phase diagram of figure~\ref{fig:hopf:lin}:
$E_0=E_1=\alpha=1$, and $\kappa=0.8$ in the isotropic limit (solid lines)
and $\epsilon_0=\mu_0=1$ for the complete set of equations (dashed lines).
The fast bursts of plastic deformation are accompanied by sudden 
jumps in the free-volume.
$\Delta$ does not play a major role for the definition of the dynamical
yield stress, as seen from the almost identical values of the maximum stress;
$\Delta$, however, strongly influences smaller values of the stress as it shifts 
upwards the unstable fixed point; 
this also modifies the periodicity of stick-slip motion. 


\section{Response to a constant stress}
In a creep test, a material is tested by forcing a constant stress.
Since $\sigma$ is constant, the elastic deformation is $\epsilon=\sigma/\mu$
at all times, and the plastic deformation is determined by the constitutive equations.


\subsection{Linear response}

\subsubsection{\label{sec:lin:iso} Isotropic limit}
A small constant stress $\sigma$ is applied at time $t=t_w$.
The plastic deformation of the material is determined 
by equations~(\ref{eqn:stzdil:iso:1}-\ref{eqn:stzdil:iso:2}),
for a fixed $\sigma$.
At large times, the rate of plastic deformation reads
$$
\dot\epsilon=\dot\gamma \simeq E_0\, \left({E_1\,t\over \kappa}\right)^{-1/\kappa}\sigma
\quad.
$$
This lead to a power law variation of the compliance, $J=\gamma/\sigma$:
$$
J(t) \propto A\;\left(t^\beta-t_w^\beta\right)
\quad.
$$
If $\kappa>1$, or $\beta>0$, the compliance diverges at long times;
if $\kappa<1$, or $\beta<0$, the compliance saturates at long times: 
$$
J(\infty) = |A|\,t_w^{-|\beta|}
\quad,
$$
the system jams. This jamming is purely entropic, and does not result
from a structural anisotropy of the force network.
Note that the asymptotic value of the compliance 
decays with the increasing age of the sample.

The validity of the linear approximation must be checked
by considering the evolution of the work of plastic deformation:
$$
\sigma\,\dot\gamma\propto t^{-{1/\kappa}}
\quad;
$$
the linear approximation is justified so long as this term is negligible,
when compared to the first term in free-volume relaxation~(\ref{eqn:stzdil:iso:2}),
which decays as $\kappa/t$.
When $\kappa<1$, the work $\sigma\,\dot\gamma$ decays faster than $\kappa/t$,
and therefore, the linear approximation is valid at all times;
when $\kappa>1$, however, this approximation fails when:
$$
\alpha\,E_0\,\left({E_1\,t\over\kappa}\right)^{-1/\kappa}\sigma^2\sim \kappa/t
\quad,
$$
or,
$$
t \sim t^{\rm n.l.} = \kappa\,\left({E_1^{1/\kappa}\over \alpha\,E_0\,\sigma^2}\right)^{\kappa\over\kappa-1}
\quad.
$$
The time $t^{\rm n.l.}$ is a typical time after which the linear 
approximation fails; this time diverges when $\sigma\to0$.
It will appear clearer in the following that, when $\kappa>1$,
free-volume presents a non-vanishing stationary value for any small applied stress.
This asymptotic value is determined by the balance between density relaxation
and shear induced dilatancy. The failure of the linear approximation, for $\kappa>1$,
is due to the 
existence of a non-zero stationary value of $\chi$ for any applied stress.

\subsubsection{\label{sec:lin:lin} General case}

In the general case, the constitutive equations 
are~(\ref{eqn:stzdil:lin:1}-\ref{eqn:stzdil:lin:3}).
The constant applied stress deforms the material elastically,
and at time $t_w^+$, no plastic rearrangement has yet occurred. 
Note that the initial value of the state variable $\Delta$ 
depends on the preparation of the sample. 
In the experiments by Cloitre {\it et al},
for example, the sample is prepared by applying a strong shear stress,
above the yield stress during a rather long time.~\cite{cloitre00}
In this case, the initial value of $\Delta$ is expected
to be non-vanishing, and this leads to their observation of strain recovery.
For the sake of generality, I will consider that the initial value of $\Delta$
is non-vanishing: $\Delta(t_w^+)=\Delta_0$.
In this section, I will assume that this value of $\Delta$ is small enough
so that the linear approximation is still valid.
When plastic rearrangements occur in the material,
they induce a structuration of the material, and 
the variable $\Delta$ starts to evolve as $\dot\Delta\simeq\dot\gamma/\epsilon_0$.

The plastic rate of deformation evolves as,
$$
\dot\gamma \simeq E_0\, \left({E_1\,t\over\kappa}\right)^{-1/\kappa}(\sigma-\Delta)
\quad,
$$
and $\Delta$ as, 
$$
\dot\Delta \simeq {E_0\over\epsilon_0}\,\left({E_1\,t\over\kappa}\right)^{-1/\kappa}
\,(\sigma-\Delta)
\quad.
$$

From this equation, the relaxation of the variable $\sigma-\Delta$ is easily
obtained:
$$
\sigma-\Delta(t) \simeq (\sigma-\Delta_0)\,\exp\left[{A''\,(t_w^\beta-t^\beta)}\right]
\quad,
$$
with
$$
A'' = {E_0\over\beta \epsilon_0}{\left({E_1\over\kappa}\right)^{-1/\kappa}}
\quad.
$$
When $\kappa>1$, $\sigma-\Delta$ undergoes a KWW relaxation towards 0:
the material relaxes towards a jammed state in which the stress is supported
by structural anisotropy.
When $\kappa<1$, $\sigma-\Delta$ saturates to a non-vanishing value: 
at long times, a fraction of the stress is supported by 
the bias between arrangements, measured by $\Delta$, but another fraction 
is supported by the loss of ergodicity resulting from the
freezing of elementary shear processes.

From equation $\dot\Delta\simeq\dot\gamma/\epsilon_0$, the compliance reads,
$$
J(t) \simeq \epsilon_0\,\left(1-{\Delta_0\over\sigma}\right)\,
(1 - \exp\left[{A''\,(t_w^\beta-t^\beta)}\right])
\quad.
$$
The deformation relaxes to a constant value, which depends on the preparation of
the sample; the relaxation is either a KWW or a power law
relaxation depending on $\kappa$.
Note that the power law variation of the compliance that was found 
in the isotropic limit is recovered when $A''\to0$;
for finite $A''$, this power law behavior shows up at short times, 
when $\Delta$ is still far from saturation.

\subsection{Jamming}

In this section I start by considering the complete set of equations~(\ref{eqn:stzdil:lin:1}-\ref{eqn:stzdil:lin:3}), 
and in particular, I recall how the dynamics of $\Delta$ lead to jamming.
Then I show how this leads to an effective equation for free-volume which
is valid in the isotropic limit or in the general case.

\begin{figure}
\narrowtext
\begin{center}
\unitlength=0.005\textwidth
\begin{picture}(110,60)(8,-8)
\put(12,40){\makebox(0,0){\large$\Delta$}}
\put(85,-13){\makebox(0,0){\large$\sigma$}}
\put(10,-13){\resizebox{90\unitlength}{!}{\includegraphics{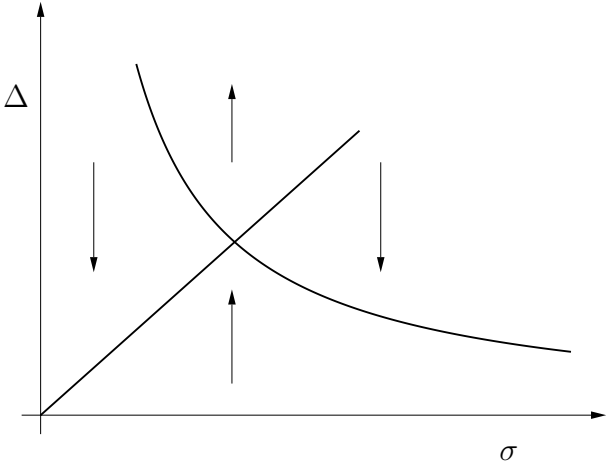}}}
\end{picture}
\end{center}
\caption{\label{fig:delta}
Bifurcation for the dynamics of the variable $\Delta$ as defined by 
equation~(\protect\ref{eqn:deltadyn}). The intersection between the two branches 
of solutions define the yield stress $\sigma_y$. 
For $\sigma<\sigma_y$, the stable solution is the jammed state, while it is the 
plastic flow for $\sigma>\sigma_y$.~\protect\cite{falk98}
}
\end{figure}

A constant stress $\sigma$ is applied. 
The dynamics of $\Delta$ is governed by the following equation:
\begin{equation}
\dot\Delta = {E_0\over\epsilon_0}\,\exp\left[-{1\over\chi}\right]\,
\left(\sigma-\Delta\right)\;\left(1-\mu_0\,\sigma\,\Delta\right)
\quad.
\label{eqn:deltadyn}
\end{equation}
Since the normalized free-volume $\chi$ enters this equation only through
a positive factor, the time-dependent value of $\chi$ does not change 
the stability analysis performed at any fixed free-volume.
The bifurcation diagram for  $\Delta$ is drawn figure~\ref{fig:delta}.

The dynamics of the variable $\Delta$ leads to jamming for any stress smaller
than the yield stress~\cite{falk98}
$$
\sigma_y = {1\over\sqrt{\mu_0}}
\quad,
$$
in which case, $\Delta = \sigma$ and $\dot\epsilon=0$.
Above the yield stress, the steady shear flow becomes a stable solution, 
in which case,
$$
\Delta = {1\over\mu_0\,\sigma}
\quad,
$$
and
\begin{equation}
\dot\epsilon = 
{E_0}\,\exp\left[-{1\over\chi}\right]\,\left(\sigma-{1\over\mu_0\,\sigma}\right)
\quad.
\label{eqn:eps:creep}
\end{equation}

If $\sigma<\sigma_y$, the systems jams, 
and the second term in the equation of motion for the free-volume vanishes.
Free-volume continues to relax logarithmically in time while the system does not deform.

When $\sigma>\sigma_y$, the system presents a non-zero plastic flow,
and the evolution of free-volume is governed by the following asymptotic equation:
\begin{equation}
\dot\chi = -E_1 \exp\left[{-{\kappa\over\chi}}\right]
+\alpha {E_0}\,\exp\left[-{1\over\chi}\right]\,\left(\sigma^2-{1\over\mu_0}\right)
\quad.
\label{eqn:chi:creep}
\end{equation}
At the level of this equation, the jamming resulting from STZ's 
shows up only through the term $1/\mu_0$ in equation~(\ref{eqn:chi:creep}), 
which shifts the square of the applied stress.
The stability analysis pertains essentially to the isotropic limit.
From this point on, I will consider the response to a constant stress 
as governed by equations~(\ref{eqn:eps:creep}) and (\ref{eqn:chi:creep})
(which amounts to neglect the influence of $\Delta$ during transients).

\subsection{Brittleness and ductility}
\begin{figure}
\narrowtext
\begin{center}
\unitlength=0.005\textwidth
\begin{picture}(110,60)(8,-8)
\put(10,40){\makebox(0,0){\large$\sigma$}}
\put(13,20){\makebox(0,0){\large$\sigma^*$}}
\put(85,-13){\makebox(0,0){\large$\chi$}}
\put(10,-13){\resizebox{90\unitlength}{!}{\includegraphics{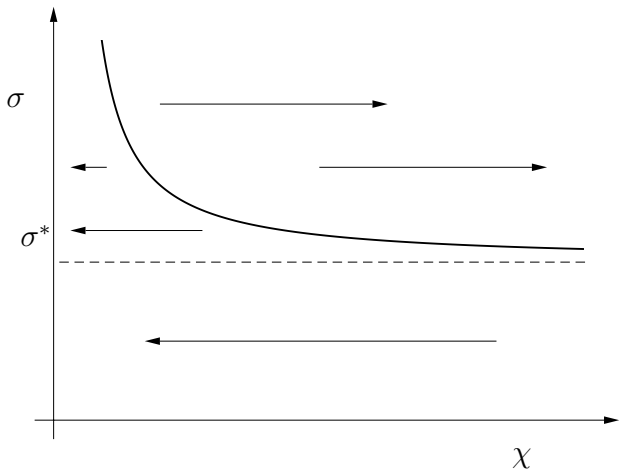}}}
\end{picture}
\end{center}
\caption{Bifurcation diagram for the free-volume dynamics defined by 
equation~(\protect\ref{eqn:chi:creep}), for $\kappa<1$. 
For all $\sigma<\sigma^*$, free-volume continues relaxation towards 0;
shear deformation occurs at a vanishingly small rate; the material creeps.
The sign of free-volume relaxation can be changed only by applying a stress at least
larger than $\sigma^*$; the required stress depends on the age of the material.
The steady solution is always unstable: large stresses lead to break up.}
\label{fig:chi:2}
\end{figure}

I now study the free-volume dynamics which result from equation~(\ref{eqn:chi:creep}).
The material dilates iff $\sigma>\sigma_y^*$, with:
$$
\sigma_y^* = \sqrt{{1\over\mu_0} +
{E_1\over\alpha \,E_0} \exp\left[{{1-\kappa\over\chi}}\right]}
$$
and contracts otherwise. The resulting bifurcation diagram for $\chi$
is drawn figure~\ref{fig:chi:1} and~\ref{fig:chi:2} for $\kappa>1$ and $\kappa<1$
respectively.

\subsubsection{Soft glasses}

For $\kappa>1$, free-volume dynamics admit a stationary solution for
any $\sigma\in[\sigma_y,\sigma^*]$, with,
$$
\sigma^* = \sqrt{{1\over\mu_0} + {E_1\over\alpha \,E_0}}
\quad.
$$
For values of the applied stress in the interval $[\sigma_y,\sigma^*]$, 
the material undergoes steady plastic deformation: it is ductile.
The shear deformation reads,
$$
\dot\epsilon = 
{E_1\over\alpha}\;
\left(\alpha\,E_0\over E_1\right)^{{\kappa\over\kappa-1}}
\;{1\over\sigma}\;\left(\sigma^2-{1\over\mu_0}\right)^{{\kappa\over\kappa-1}}
\quad,
$$
and the strain rate behaves as a power of the stress for large stresses.
When the applied stress varies in the interval, $\sigma\in[\sigma_y,\sigma^*]$,
the strain rate can display a whole interval of values
$$
\dot\epsilon\in\left[0,E_0\,\left(\sigma^*-{1\over\mu_0\,\sigma^*}\right)\right]
\quad.
$$
For larger values of $\sigma$, above $\sigma^*$, free-volume diverges, leading
to the break-up of the material.

\begin{figure}
\narrowtext
\begin{center}
\unitlength=0.005\textwidth
\begin{picture}(110,60)(8,-8)
\put(10,48){\makebox(0,0){\large$\sigma$}}
\put(13,40){\makebox(0,0){\large$\sigma^*$}}
\put(13,-1){\makebox(0,0){\large${1\over\mu_0}$}}
\put(85,-13){\makebox(0,0){\large$\chi$}}
\put(10,-13){\resizebox{90\unitlength}{!}{\includegraphics{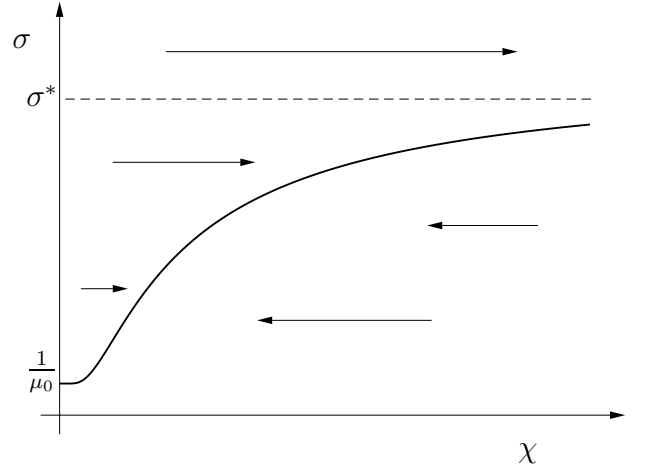}}}
\end{picture}
\end{center}
\caption{Bifurcation diagram for the free-volume dynamics defined by 
equation~(\protect\ref{eqn:chi:creep}), for $\kappa>1$. 
For any value of the stress
in the interval $[\sigma_y,\sigma^*]$, there exist a stationary value of $\chi$. 
The applied stress stops aging, and the material undergoes steady 
plastic deformation.
Above $\sigma^*$, free-volume diverges leading to the break-up of the material.
}
\label{fig:chi:1}
\end{figure}

\begin{figure}
\narrowtext
\begin{center}
\unitlength = 0.005\textwidth
\begin{picture}(100,110)(5,0)
\put(10,5){\resizebox{90\unitlength}{!}{\includegraphics{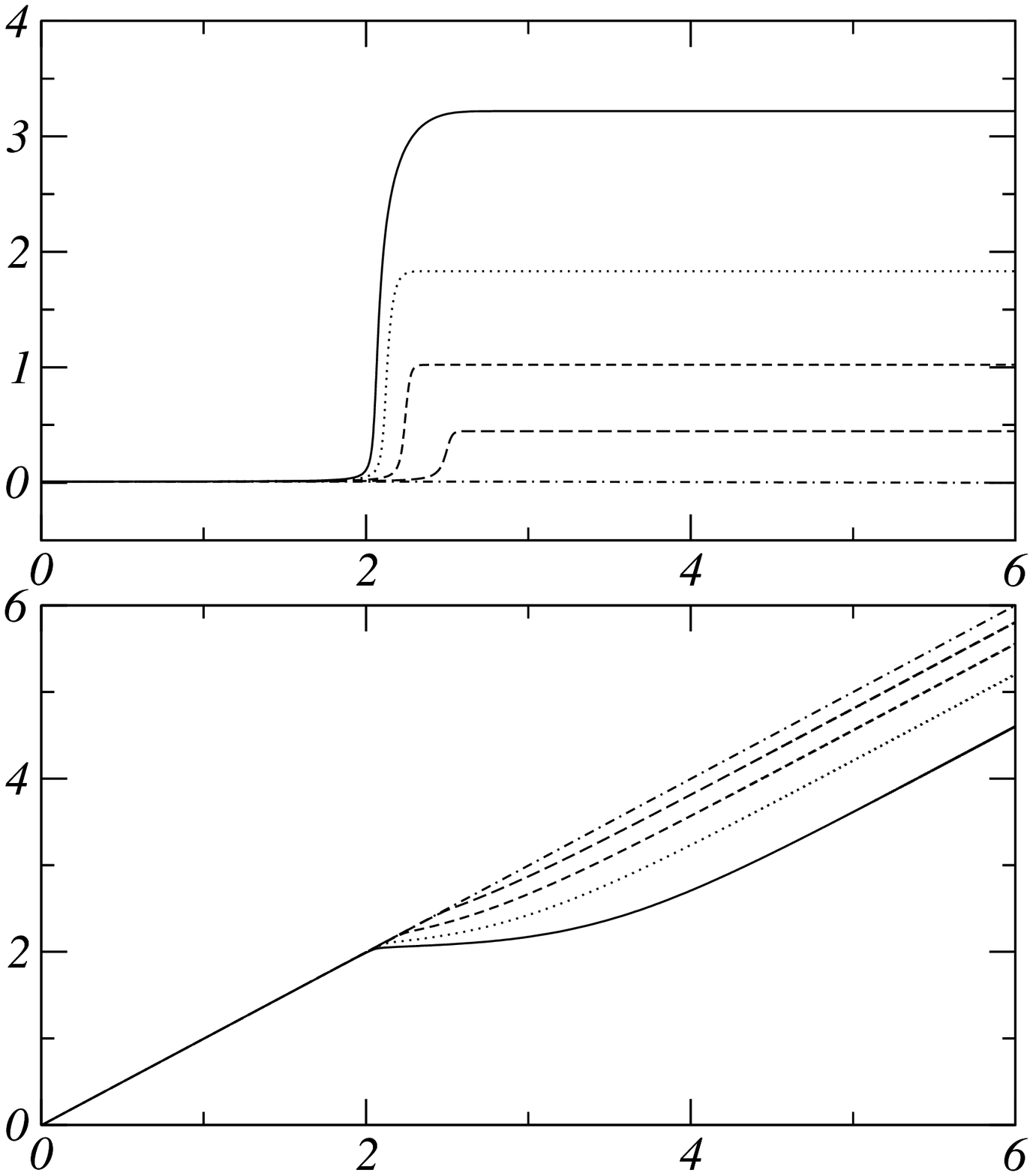}}}
\put(5,95){\makebox(0,0){\large $1/\chi$}}
\put(5,50){\makebox(0,0){\large $\log_{10}(J)$}}
\put(90,2){\makebox(0,0){\large $\log_{10}(t)$}}
\end{picture}
\end{center}
\caption{
Numerical integration of equations~(\ref{eqn:stzdil:iso:1}) 
and~(\ref{eqn:stzdil:iso:2}) for constant stresses $\sigma$, and for $\kappa>1$.
Parameters are $E_0=E_1=\alpha=\mu=1$, and $\kappa=2$;
for these parameter, $\sigma^*=1$.
Initial value of the free-volume is $\chi=100$ and $t_w=0$.
Top: $1/\chi$ as a function of time for 
$\sigma = 0.2, 0.4, 0.6, 0.8, 1$ from top to bottom; for $\sigma=1$, and for
all values above, free-volume diverges.
Bottom: $\log_{10}(J)$ as a function of time for $\sigma$ 
increasing from bottom to top.
}
\label{fig:compl:soft}
\end{figure}

The resulting dynamics for the compliance and for the free-volume are presented figure~\ref{fig:compl:soft} for different values of the driving stress $\sigma$.
The first part of the dynamics is determined by the initial value of 
free-volume and the corresponding timescale.
As easily seen on the bifurcation diagram~(\ref{fig:chi:1}), for small stresses, 
an initially large free-volume decays, and slows down the plastic deformation. 
This results in a plateau for the dynamics of the compliance; a larger times, free-volume converges 
towards a constant value, and the dynamics of $J$ resumes a steady increase.

These curves are to compare with measurements of particle diffusion:
the particle diffusion in a dense material is expected to be dominated 
by collective processes; the diffusive constant is then $\propto\exp[-1/\chi]$, 
and the average motion of a particle in a glassy medium is essentially
governed by the same equation as $J$. 
The features displayed of figure~\ref{fig:compl:soft}
are amazingly similar to the curves found in 
several recent experiments.~\cite{megen98,cloitre00}

If free-volume diverges, just before break-up, 
the current equations indicate
that the strain rate should display a behavior of the form,
$$
\dot\epsilon = E_0\,\left(\sigma-{1\over\mu_0\,\sigma}\right)
\quad.
$$
However, this expression must be taken with caution,
since the strongly out-of-equilibrium behavior of a material during break-up, 
when $\chi$ is large, is beyond the scope of the current theory.
In particular, assumption has been made that the distribution 
of voids is close to an equilibrium state.
During break-up, the distribution of free-volume is expected to display strong
heterogeneities leading to the nucleation of fractures.

\subsubsection{Hard glasses}

For $\kappa<1$, and for any $\sigma<\sigma^*$, $\dot\chi<0$: 
free-volume relaxes to 0. In this case, the applied stress does not stop aging,
shear deformation is vanishingly small: the material creeps.
At long time, the linear term dominates free-volume relaxation: the calculation
of linear response that has been presented in section~\ref{sec:lin:iso} is valid,
and indicates that compliance saturates: the system jams.
In general, part of this jamming is supported by the structuration of the material,
which lowers the value of the stress that enters equation~(\ref{eqn:chi:creep});
the remaining stress is supported by entropic freezing of the rearrangement dynamics.

The sign of free-volume relaxation can be inverted only for larger stresses, 
above $\sigma_y^*(\chi)$; it this case, the response depends sensitively on 
the value of $\sigma$ and on the state of the material at the time $t_w$
when stress is applied:
a larger stress is required to break an older material.
The stationary state is unstable: when $\sigma$ is large enough 
to invert the sign of free-volume dynamics, the material breaks up.
In this case, there are two types of response: creep for low stress, and break-up
for large values of the stress. The material is brittle.

\begin{figure}
\narrowtext
\begin{center}
\unitlength = 0.005\textwidth
\begin{picture}(100,110)(5,0)
\put(10,5){\resizebox{90\unitlength}{!}{\includegraphics{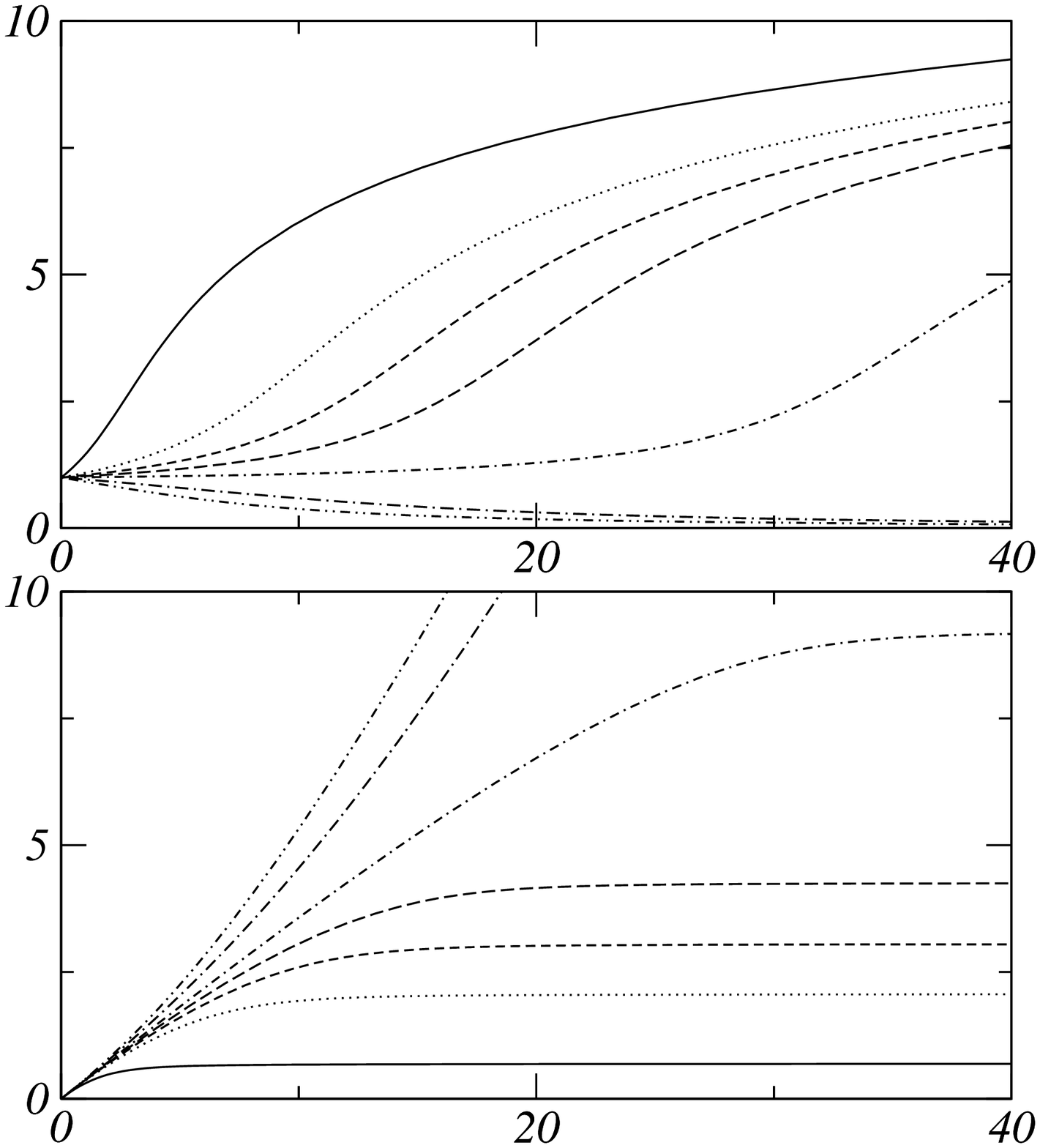}}}
\put(5,95){\makebox(0,0){\large $1/\chi$}}
\put(5,50){\makebox(0,0){\large $J$}}
\put(90,2){\makebox(0,0){\large $t$}}
\end{picture}
\end{center}
\caption{
Numerical integration of equations~(\ref{eqn:stzdil:iso:1}) 
and~(\ref{eqn:stzdil:iso:2}) for constant stresses $\sigma$.
Parameters are $E_0=E_1=\alpha=\mu=1$, and $\kappa=0.8$.
Initial value of the free-volume is $\chi=1$ and $t_w=0$.
For these values of the parameters, and of initial conditions, 
the critical value of $\sigma$ where the free-volume diverges is, 
$\sigma_y^*\simeq 1.105$.
Top: $1/\chi$ as a function of time for 
$\sigma = 0.5, 1, 1.05, 1.075, 1.1, 1.15, 1.2$ from top to bottom.
Bottom: $J$ as a function of time for $\sigma$ increasing from bottom to top.
}
\label{fig:compl:hard}
\end{figure}

The resulting dynamics for free-volume and for the compliance are 
displayed figure~\ref{fig:compl:hard}. The saturation of the compliance
is clearly seen for small stresses while the divergence occurs at 
a value at $\sigma=\sigma_y^*(\chi)$.
The shape of these curves changes drastically in a very small interval of $\sigma$
around $\sigma_y^*(\chi)$ because of long transient to escape from 
the unstable fixed point. These features are commonly observed in creep tests.~\cite{hassan95,larson99}

\subsubsection{Jamming and freezing}

Materials with $\kappa>1$ are liable to display ductile behavior,
but can break; those with $\kappa<1$ either creep or break-up, 
and could be classified as unconditionally brittle. 
The study of STZ equations one a one-dimensional
elasto-plastic decohesion has shown the importance of dynamics in ductile to brittle
transition.~\cite{lobkovsky98}
The current analysis slightly contradicts those results,
and indicates that two major types of materials
can be identified, which display different qualitative behavior: 
this is in agreement
with the common idea that some materials are ductile, some other brittle.

In fact, for hard materials ($\kappa<1$), the existence of a strong entropic crisis
precludes dynamics of rearrangements to come into play: intrinsic
properties of the material completely dominate the dynamical process of deformation.
The existence of a dynamically driven transition between ductile and brittle
behavior is still expected for soft materials ($\kappa>1$) for which the dynamics
or arrangements can fully play its role.
The interplay between dynamics and intrinsic glassy properties of material, 
as displayed in the current approach, seems a promising perspective.

As far as fracture mechanics is concerned, this picture indicates that
during the breaking of a hard glass, the motion of molecules in a material
should be essentially localized around the surface of a crack,
where there is some free-volume,
whereas in the bulk, plastic deformation can occur only in places where a very large
stress in concentrated. This is apparently related to a 
theoretical approach of fracture mechanics,
recently introduced by Brener and Spatschek,~\cite{brener02} 
where the dynamics of the fracture is controlled by diffusion 
of molecules along the surface.
For soft glassy materials, on the contrary, the propagation
of a fracture is expected to induce more easily observable 
plastic deformation in the bulk of the material.



\section{Conclusion}

The current work relies on the introduction of dynamical equations for 
the intensive quantity associated with entropic fluctuations 
of molecular configurations.
In this work, this quantity is identified to free-volume,
but other physical pictures could be given.
For example, free-volume is related to the average number of contacts per molecule;
in a foam, the deformation of the bubbles can be accounted for a effective 
free-volume that is determined by the packing of effective overlapping
particles as those used in the simulations by Durian,~\cite{durian95,durian97} 
even though the relative proportions
of a fluid mixture is kept constant; in a colloidal suspension, the existence
of electrostatic screening contributes to an effective size of the particle
that depends on the packing structure. 
In fact, the assumption made in this work is rather general,
and many situation can be expected to lead to a dynamically 
driven effective free-volume; this may explain the ubiquity of glassy behavior.


The theoretical approach to glassy materials can be summarized as:
how ergodicity breaks? or what ergodicity breaking is sufficient to account
for slow modes of relaxation. The introduction of a whole distribution of timescales
is a very general way to account for non-ergodicity.
Here I assume that a much weaker ergodicity breaking is sufficient,
and this is directly supported by the observation that fluctuation-dissipation
theorem holds in a weak (but, not so weak) form.
This indicates that the underlying
distributions of state variables should not (in fact) be too weird,
and that an effective width accounts for its essential features.
However, since the system is out-of-equilibrium, 
this width must, at least, become a dynamical quantity.
The equation proposed appear as a very minimal first order closure 
of the dynamics of an underlying distribution of voids.

The set of curves displayed in figure~\ref{fig:relax3} 
is very reminiscent of the temperature
dependency of relaxation spectra.~\cite{larson99} Obviously, the parameters
of the theory should depend on temperature and other thermodynamical parameters, 
but here I purposely do not intend to address this question. 
There are (at least) two reasons for that: firstly, because
glassy behavior is observed in many systems, 
in particular in suspensions or in granular materials, 
where thermodynamical temperature is not a relevant parameter.
Secondly, because it seems to me more interesting to extract all possible conclusions
from the simple equations proposed without trying to incorporate them into a more
specific setup. I do think that an effective dynamically driven free-volume
should emerge from microscopic dynamics under minimal assumptions. 
Therefore, it seems to me important to try to get
some fundamental understanding of the consequences of such assumptions for a fixed
set of parameters, without complicating the discussion.
The dependency of the parameters of this theory on thermodynamic quantities
is definitely an important issue to address in the future, but it first requires
to have some intuition about the outcome of those equations. 
This work is intended to provide a grid of possible behavior displayed
by simple equations, and I hope it can be helpful for further developments.

Finally, one major interest of the current work is, in my opinion, that is gives 
a unified picture where different mechanisms for jamming coexist.
This coexistence of jamming with non-linear rheology has been observed 
experimentally,
but could not be captured by earlier model of glassy rheology.
Moreover, it opens the way towards a description of glassy materials 
where thermodynamics (that determine the parameters of constitutive equations)
and dynamics coexists. 

\acknowledgements

This work has primarily benefited from the support and encouragements of
Jean Carlson and Jim Langer who received me among their people
in the busy heights of UCSB;
I acknowledge the opportunity they gave me to pursue my ideas,
and their continuous interest for my work.
I have enjoyed useful discussions with them, and also with others:
Ralph Archuleta, Pascal Favreau, Anthony Foglia, Delphine Gourdon,
Jacob Israelachvili, Daniel Lavall\'ee, Jean-Christophe Nave.
I also would like to thank Daniel Bonn for giving me early versions of his papers,
which were helpful in the writing of this manuscript.

This work was supported by the W. M. Keck Foundation,
and the NSF Grant No. DMR-9813752,
and EPRI/DoD through the Program on Interactive Complex Networks.

\thanks

\appendix
\section{\label{app:hopf} Hopf bifurcation at constant strain rate}

\subsection{Isotropic limit}
In order to carry out the stability analysis of the 
system~(\ref{eqn:sigma:0}),~(\ref{eqn:stzdil:iso:1}),~(\ref{eqn:stzdil:iso:2}),
it is convenient to introduce
the variable,
$$
\phi = \exp\left[-{1\over\chi}\right]
\quad.
$$

The system~(\ref{eqn:sigma:0}),~(\ref{eqn:stzdil:iso:1}),~(\ref{eqn:stzdil:iso:2})
reduces to two equations for $\sigma$ and $\chi$,
\begin{eqnarray*}
\dot\sigma &=& \mu\,\dot\epsilon-\mu\,E_0\,\exp\left[-{1\over\chi}\right]\sigma\\
\dot\chi &=& -E_1 \exp\left[-{\kappa\over\chi}\right] + \alpha\,E_0\,\exp\left[-{1\over\chi}\right]\,\sigma^2
\end{eqnarray*}
and the jacobian of this dynamical system reads:
$$
\left(
\matrix{
-\mu\,E_0\,\phi  & 
-\mu\,E_0\,\sigma\,\phi\,\log[\phi]^2\cr
2\,E_0\,\alpha\,\sigma\,\phi& 
\left(-\kappa\,E_1\phi^\kappa + E_0\,\alpha \,{\sigma }^2\,\phi\right)\log[\phi]^2\cr
}
\right)
\quad.
$$

In the steady state, the variable $\sigma$ verifies,
$$
\sigma^2 = {E_1\over\alpha\,E_0}\,\phi^{\kappa-1}
$$
and $\phi$ takes on the value,
$$
\phi_{\rm s}=\left({\alpha\,\dot\epsilon^2\over E_0\,E_1}\right)^{1\over\kappa+1}
\quad.
$$
For all $\kappa>0$, $\phi_{\rm s}$ is a strictly increasing function of $\dot\epsilon$.
In terms of $\phi_{\rm s}(\dot\epsilon)$, the eigenvalues $\lambda$ of the jacobian verify:
\begin{eqnarray*}
\lambda^2 +
\lambda\;\left(
\mu\,E_0\,\phi_{\rm s}
+ {E_1\,(\kappa-1)}\,\phi_{\rm s}^\kappa
\;\ln\left[\phi_{\rm s}\right]^2
\right)
&+&\\
{\mu\,E_0\,E_1\,(\kappa+1)}\;\phi_{\rm s}^{\kappa+1}\,\ln\left[\phi_{\rm s}\right]^2
&=&0
\quad,
\end{eqnarray*}
and the Hopf bifurcation is determined by a critical value of $\mu$ as a function
of $\phi_{\rm s}$:
$$
\mu_{\rm hopf} = {E_1\over E_0}\,(1-\kappa)\,\phi_{\rm s}^{\kappa-1}\,\ln[\phi_{\rm s}]^2
$$
or,
$$
\mu_{\rm hopf} = {E_1\over E_0}\,{1-\kappa\over(\kappa+1)^2}\;
\left({\alpha\,\dot\epsilon^2\over E_0\,E_1}\right)^{{\kappa-1\over\kappa+1}}\;
\ln\left[{\alpha\,\dot\epsilon^2\over E_0\,E_1}\right]^2
\quad.
$$

\subsection{General case}
In the general case, the system~(\ref{eqn:sigma:0}),~(\ref{eqn:stzdil:lin:1}-\ref{eqn:stzdil:lin:3})
involves one additional variable ($\Delta$), and the bifurcation analysis is somewhat more complicated.
In the steady state, the relation between strain rate and free-volume can be rewritten as,
$$
\dot\epsilon = {
\sqrt{E_0\,\mu_0}\,E_1\;\phi^{{\kappa/2}}
\over
\sqrt{\alpha}\;
\sqrt{E_0\alpha\,\phi^{-\kappa} 
		    + E_1\mu_0\,\phi^{-1}}
}
\quad.
$$
Since this expression is an strictly increasing function of $\phi$, 
it is possible to carry out the analysis in terms of $\phi=\phi_{\rm s}$ as an effective
parameter. When the strain rate varies between $0$ and $\dot\epsilon^*$,
$\phi$ varies on the interval $[0,1]$. The values above $1$ are meaningless
because there is no solution to the equations of motion.

The Hopf criterion can be written in the form, 
$$
{\cal H}=A\,\mu^2+B\,\mu+C=0
$$
with:
\begin{eqnarray*}
A &=&  {E_0}^2\,\alpha \,\epsilon_0\,{\phi }^2\,
\Big( \mu_0\,\left( 2\,E_0\,\alpha \,\phi  + 
        E_1\,\mu_0\,{\phi }^{\kappa } \right)  \Big.\\
&+& \Big.
     \alpha \,\epsilon_0\,\left( E_0\,\alpha \,\kappa \,\phi  + 
        E_1\,\left( 1 + \kappa  \right) \,\mu_0\,{\phi }^{\kappa } \right) \,
      {\log (\phi )}^2 \Big)\\
B &=& E_0\,E_1\,{\phi }^{1 + \kappa }\,
   \Big( {\mu_0}^2\,
\left( 2\,E_0\,\alpha \,\phi+ E_1\,\mu_0\,{\phi }^{\kappa } \right)
+ \alpha \,\epsilon_0\,\mu_0\,\Big.\\
\Big.&&\times \left( 
	E_0\,\alpha \,\left( 2\,\kappa-3  \right) \,\phi +
 	2\,E_1\,\left( \kappa -1 \right) \,\mu_0\,{\phi }^{\kappa } 
	\right)\, {\log (\phi )}^2 \Big.\\
\Big.&+&
{\alpha }^2\,{\epsilon_0}^2\,\left( \kappa -1  \right) \,
      	\left( E_0\,\alpha \,\kappa \,\phi  + 
        E_1\,\left( 1 + \kappa  \right) \,\mu_0\,{\phi }^{\kappa } \right) \,
      {\log (\phi )}^4 \Big)\\
C &=& {E_1}^2\,\left(\kappa  -1  \right) \,\mu_0\,{\phi }^{2\,\kappa }\,
   \left( E_0\,\alpha \,\phi  + E_1\,\mu_0\,{\phi }^{\kappa } \right) \,
   {\log (\phi )}^2\, \\
&&\times\left( \mu_0 +  \alpha \,\epsilon_0\,\left(\kappa -1 \right) \,{\log (\phi )}^2 \right)
\end{eqnarray*}
The quantity ${\cal H}$ has the opposite sign of the real part 
of the complex eigenvalues of the jacobian: the steady state is stable,
iff ${\cal H}>0$.

For $\kappa>1$, and for any $\phi\in[0,1]$, $C$ is positive: the two values of $\mu$, 
solutions of ${\cal H}=0$ have the same sign. This sign can change only
at a point where $C$ vanishes, that is $\phi=0$ or $1$. The sign shared by these
solutions on the interval $\phi\in[0,1]$ can then be evaluated by taking the limit
value of the coefficient $B$ close to 0 or 1: it is positive.
Therefore, the equation ${\cal H}=0$ has no real positive solution for $\kappa>1$:
there is no Hopf instability. It can be checked that there is no other instability,
therefore, for $\kappa>1$, the system is stable.

I am now considering the case when $\kappa<1$.
To get a first idea of the phase diagram, in the parameter space $\{\mu,\phi\}$,
the quantity ${\cal H}$ can be evaluated on the line $\mu=0$: ${\cal H}(\mu=0,\phi)=C$.
On this line, the positiveness of ${\cal H}$ is then equivalent to,
$$
{\log (\phi )}^2 > {\mu_0\over\alpha \,\epsilon_0\,\left(1-\kappa\right) }
\quad,
$$
which requires that:
$$
\phi < \phi_{\rm hopf}^-(\mu=0) = 
\exp\left[-\sqrt{\mu_0\over\alpha \,\epsilon_0\,\left( 1-\kappa\right) }\right]
\quad.
$$

For $\phi>\phi_{\rm hopf}^-$ the coefficient $C$ is negative, hence, equation ${\cal H}=0$ 
has a single positive solution.
On the interval $[0,\phi_{\rm hopf}^-]$ this equation admits two solutions which are positive
because $B$ is negative close to $\phi=0$: the solution $\mu_{\rm hopf}^-$ which vanishes
at the point $\phi_{\rm hopf}^-$ and the solution $\mu_{\rm hopf}^+$ which exists on the whole
interval $[0,1]$. The unstability occurs for $\mu\in[\mu_{\rm hopf}^-,\mu_{\rm hopf}^+]$.
At small shear rates, or small $\phi$, the criterion ${\cal H}$ is dominated by:
\begin{eqnarray*}
A &\simeq&  E_0^2\,E_1\,\alpha^2\,\epsilon_0^2\,\mu_0\,
\left( 1 + \kappa  \right) \,{\phi }^{2+\kappa }\,{\log (\phi )}^2\\
B &\simeq& E_0\,E_1^2\,{\alpha }^2\,{\epsilon_0}^2\,\mu_0\,\left(\kappa^2-1\right)\,
{\phi }^{1+2\kappa } \, {\log (\phi )}^4\\
C &\simeq& {E_1}^3\,\alpha \,\epsilon_0\,\mu_0^2\,\left(\kappa-1\right)^2\,
{\phi }^{3\,\kappa }\,{\log (\phi )}^4
\end{eqnarray*}
whence,
\begin{eqnarray*}
{B\over A} &\simeq& {E_1\over E_0}\,(\kappa-1)
{\phi }^{\kappa-1} \, {\log (\phi )}^2\\
{C\over A} &\simeq& {E_1\,\mu_0\over E_0^2\,\alpha\,\epsilon_0}\,
{\left(\kappa-1\right)^2\over\kappa+1}\,
{\phi }^{2\,\kappa-2}\,{\log (\phi )}^2\\
{C\over B} &\simeq& {\mu_0\over E_0\,\alpha\,\epsilon_0}\,
{\kappa-1\over\kappa+1}\,{\phi }^{\kappa-1}
\end{eqnarray*}
The solution $\mu_{\rm hopf}^+$ can then be identified to:
$$
\mu_{\rm hopf}^+ \simeq -{B\over A}\simeq 
{E_1\over E_0}\,(1-\kappa) {\phi }^{\kappa-1} \, {\log (\phi )}^2
\quad,
$$
close to $\phi=0$.
The same expression was found in the isotropic limit, 
but here, the relation between $\phi$ and the shear rate, 
$\dot\epsilon$ is different.
The solution $\mu_{\rm hopf}^-$ can be expanded as,
$$
\mu_{\rm hopf}^- \simeq -{C\over B}\simeq  
{\mu_0\over E_0\,\alpha\,\epsilon_0}\,
{1-\kappa\over\kappa+1}\,{\phi }^{\kappa-1}
$$

\end{document}